\documentclass[letterpaper,twocolumn,10pt]{article}
\usepackage{style}

\usepackage{multirow}
\usepackage{amsmath}
\usepackage{amssymb}
\usepackage{mathtools}
\usepackage{amsthm}
\usepackage{xspace}
\usepackage[capitalize,noabbrev]{cleveref}
\usepackage{amsfonts}  
\usepackage{algorithm}  
\usepackage{algpseudocode} 
\theoremstyle{plain}

\theoremstyle{definition}

\theoremstyle{remark}

\usepackage[textsize=tiny]{todonotes}
\usepackage{subcaption} 

\begin{document}
\date{}

\title{\bf FaceSwapGuard: Safeguarding Facial Privacy from DeepFake Threats through Identity Obfuscation}

\author{
Li Wang\textsuperscript{1}\ \ \
Zheng Li\textsuperscript{1}\ \ \
Xuhong Zhang\textsuperscript{2}\ \ \
Shouling Ji\textsuperscript{2}\ \ \
Shanqing Guo\textsuperscript{1}
\\
\\
\textsuperscript{1}\textit{Shandong University} \ \ \ 
\textsuperscript{2}\textit{Zhejiang University} \ \ \
}

\maketitle

\begin{abstract}
DeepFakes pose a significant threat to our society.
One representative DeepFake application is face-swapping, which replaces the identity in a facial image with that of a victim. 
Although existing methods partially mitigate these risks by degrading the quality of swapped images, they often fail to disrupt the identity transformation effectively.
To fill this gap, we propose FaceSwapGuard (\texttt{FSG}), a novel black-box defense mechanism against deepfake face-swapping threats. 
Specifically, \texttt{FSG} introduces imperceptible perturbations to a user’s facial image, disrupting the features extracted by identity encoders. When shared online, these perturbed images mislead face-swapping techniques, causing them to generate facial images with identities significantly different from the original user.  
Extensive experiments demonstrate the effectiveness of \texttt{FSG} against multiple face-swapping techniques, reducing the face match rate from 90\% (without defense) to below 10\%. 
Both qualitative and quantitative studies further confirm its ability to confuse human perception, highlighting its practical utility. 
Additionally, we investigate key factors that may influence FSG and evaluate its robustness against various adaptive adversaries.
\end{abstract}

\section{Introduction}

\begin{figure}[t]
\centering
\includegraphics[width=\columnwidth]{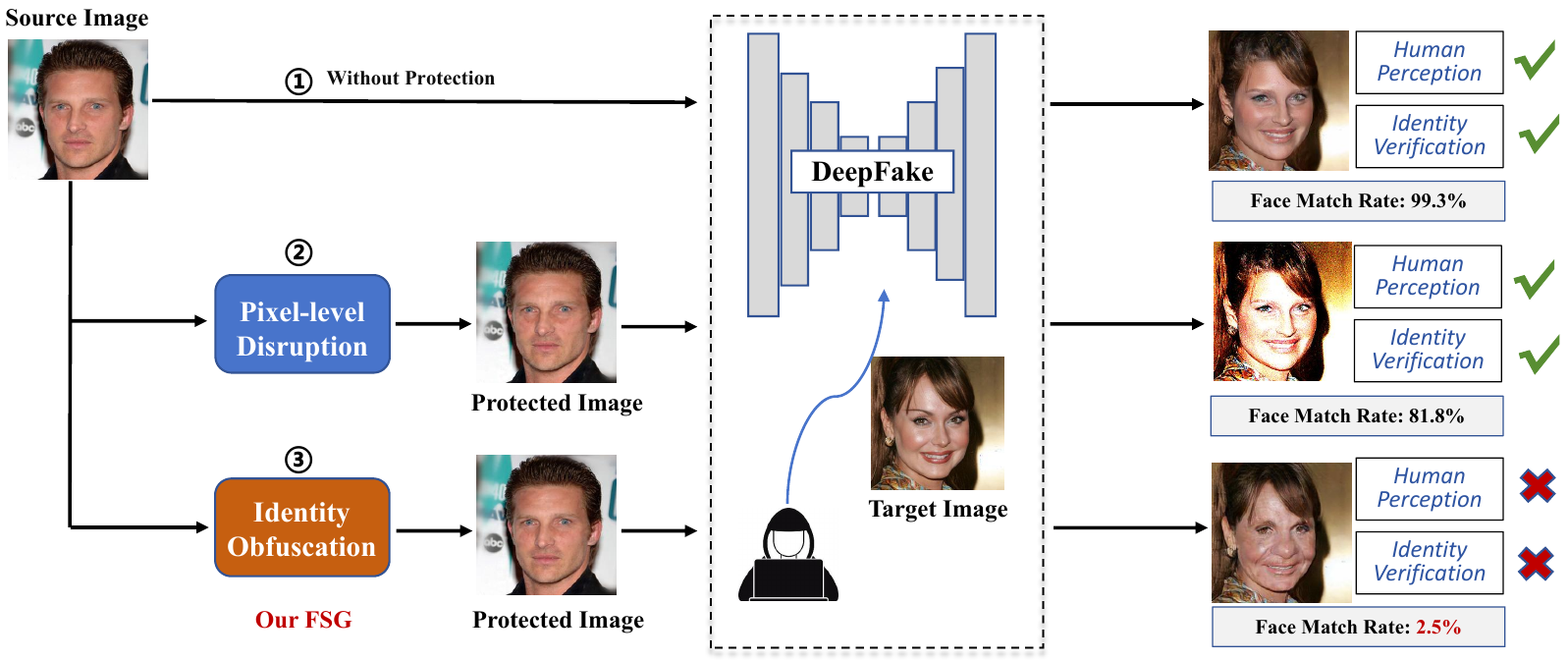} 
\caption{A high-level illustration of \texttt{FSG}. The source image depicts the victim, while the DeepFake replaces the target image's identity with the victim's.} 
\label{overview1}
\end{figure}

Advances in generative AI ~\cite{jovanovic2022generative} have led to highly realistic generated images, marking the rise of the DeepFake era~\cite{verdoliva2020media}. 
A notable example is face swapping~\cite{chen2020simswap,wang2024deepfaker}, which replaces the identity in a target image with a victim's while maintaining pose, expression, and background (See \autoref{overview1}). 
It enables identity theft~\cite{korshunov2018deepfakes,tariq2022real} by bypassing facial recognition systems and spreading misinformation, damaging reputations~\cite{meng2023ava}, thereby escalating societal and security threats.  
Besides, Li \textit{et al.}~\cite{li2022seeing} show that it can also evade liveness detection, circumventing facial verification systems used in security-critical applications. 
Quantitatively, on Face++'s recognition API, the Face Match Rate (FMR)—measuring the likelihood that two facial images belong to the same person—reaches 99.3\% between 1,000 generated images and user images, highlighting significant security risks. To address these threats, obfuscating identities in generated images is crucial. This ensures that even if a user's image is misused, the resulting DeepFake cannot be recognized as the user's true identity, effectively protecting facial privacy.

Many existing methods defend against GAN-based facial manipulation by adding adversarial perturbations to victim images, aiming to degrade the quality of generated images~\cite{ruiz2020disrupting,yeh2020disrupting}. 
However, these methods focus on maximizing pixel-level distortion between adversarial and original generated images. 
In black-box scenarios, where DeepFake models are unknown, they often only introduce blurriness or artifacts~\cite{dong2023restricted}, as shown in \autoref{overview1}. 
As a result, they fail to disrupt identity preservation during generation. 
While image quality decreases, the subject's identity remains intact, potentially allowing the images to bypass identity verification and human perception. 
Additionally, these methods rely on specific generative models for perturbation, limiting their transferability and efficiency across different face-swapping models.

To address this gap, we propose \texttt{FSG}, a novel defense approach that mitigates DeepFake face-swapping threats through identity obfuscation. A high-level overview of \texttt{FSG} is shown in \autoref{overview1}. Our approach is inspired by the pipeline of advanced feature-level face-swapping techniques, which use an identity encoder to extract and preserve features from source images during injection and decoding, ensuring identity preservation in generated images. 
To disrupt this process, \texttt{FSG} introduces imperceptible perturbations to victim images, altering the features extracted by the identity encoder. This causes confusion in both identity and visual perception of the generated images. Importantly, \texttt{FSG} computes these protected images in a model-agnostic manner, ensuring broad applicability. 
Specifically, we construct a surrogate model for the identity encoder and add perturbations to maximize identity deviation between protected images and user source images. 
To enhance the perceptual difference between generated and source images during decoding, we constrain the loss of the last $k$ layers, which capture higher-level semantics and identity-related features. 
Additionally, we improve robustness by incorporating random image transformations (e.g., resizing and cropping) during optimization, simulating common DeepFake operations. 
As a result, our method ensures that face-swapped images not only fail identity verification but also visually diverge from the user’s genuine identity.

Extensive experiments demonstrate \texttt{FSG}'s effectiveness across various face-swapping models in black-box scenarios. 
First, we assess \texttt{FSG}'s identity obfuscation performance by measuring the FMR on academic facial recognition models and commercial APIs. 
Results show a significant FMR drop between generated and source images, reducing FMR from over 90\% to below 10\% on multiple APIs. 
We also provide perceptual similarity metrics and visualizations to confirm its effectiveness in confusing human perception. Ablation experiments highlight the critical role of each \texttt{FSG} component. 
Additionally, protected images remain robust against potential adaptive attacks. 
Finally, we evaluate \texttt{FSG} on other DeepFake types, including diffusion-based models (see Appendix \ref{A6} for details).

The main contributions are summarized as follows:
\begin{itemize}
    \item We propose FaceSwapGuard (\texttt{FSG}), a novel proactive defense method to safeguard facial images against DeepFake face-swapping. By adding imperceptible perturbations to user photos, \texttt{FSG} generates protected images that effectively obscure identity features in face-swapped outputs.
    
    \item We design an alternative identity encoder to compute protected images, offering a flexible solution independent of generative models. Additionally, we employ intermediate feature map loss and differentiable random image transformations to enhance visual confusion and transferability.
  
    \item Extensive evaluations on industrial face verification APIs and academic recognition models demonstrate \texttt{FSG}'s effectiveness in identity obfuscation. Quantitative and qualitative results confirm its ability to confuse visual perception. We also verify the robustness of protected images against potential adaptive attacks. Additionally, \texttt{FSG} shows promise in generalizing to other DeepFake models, including diffusion-based approaches.
\end{itemize}

\section{Related Work}

\subsection{Face Swapping}
Face swapping employs advanced AI to manipulate visual content, altering identity features and generating face-swapped images with specific expressions and movements~\cite{mirsky2021creation}. This field has seen extensive research, with numerous methods proposed~\cite{tolosana2020deepfakes}. Early image-level methods transfer target face attributes to the source face using segmentation masks for blending~\cite{nirkin2018face}. For example, FSGAN~\cite{nirkin2019fsgan} uses a reenactment network to transfer expressions and poses, followed by a blending network combining Poisson optimization ~\cite{perez2003poisson} and perceptual loss. However, these methods are sensitive to source images and often fail to preserve target image attributes, leading to artifacts. 
State-of-the-art feature-level methods extract identity features from the source face, and attribute features from the target face ~\cite{bao2018IPGAN,chen2020simswap}, then decode these features into generated images. FaceShifter~\cite{li2019FaceShifter} produces high-fidelity results by adaptively integrating target attributes (e.g., expression, lighting) and handling occlusions in a self-supervised manner. SimSwap~\cite{chen2020simswap} further improves this by introducing a weak feature matching loss, preserving attributes, and avoiding mismatched poses and expressions. Overall, face swapping continues to evolve, focusing on reducing artifacts and improving identity and attribute preservation.

However, the malicious exploitation of face-swapping technology could pose significant threats to security and privacy ~\cite{tan2024rethinking}, particularly by presenting unprecedented challenges to identity verification services that rely on underlying facial recognition technology.
Tariq \textit{et al.} ~\cite{tariq2022real} examine the vulnerability of the celebrity recognition APIs to DeepFake attacks, revealing significant weaknesses in identity verification systems.
Li \textit{et al.} ~\cite{li2022seeing} investigate the security of facial liveness verification systems, including facial recognition technology, highlighting vulnerabilities in widely deployed APIs within the evolving attack-defense landscape.

\subsection{Defense Method}
To combat face-swapping threats, researchers have focused on detection~\cite{gu2022delving, wang2024deepfake}, primarily using binary classification models~\cite{masood2023trend1}.  
However, these passive methods detect DeepFakes only after creation, allowing security breaches, identity theft, or reputational damage to have already occurred first.

Recent efforts in proactive defense against facial manipulation involve adding imperceptible adversarial perturbations to user images to disrupt the synthesis of generated images ~\cite{ruiz2020disrupting, huang2021initiative}. For instance, Yeh \textit{et al.} \cite{yeh2020disrupting} propose distortion and nullifying attacks, which maximize the distance between adversarial and original outputs or minimize the distance between adversarial images and inputs, respectively. These methods effectively protect against image-translation-based DeepFake attacks in white-box scenarios. In black-box settings, Ruiz \textit{et al.} ~\cite{ruiz2020protecting} propose a query-based adversarial attack, though its practicality is limited due to high query demands.
Existing defenses mainly target attribute editing or facial reenactment (e.g., StarGAN ~\cite{choi2018stargan}, GANimation ~\cite{pumarola2018ganimation}), neglecting face-swapping models that manipulate hard biometric features, which pose greater threats. Dong \textit{et al.} \cite{dong2023restricted} use adversarial example transferability to resist face swapping in restricted black-box scenarios. However, like prior methods, they disrupt images at the pixel level, reducing visual quality without significantly altering identity features, thus failing to fully mitigate identity theft risks. Additionally, current methods often rely on specific DeepFake models, limiting their effectiveness across different models.

\section{Methodology of FaceSwapGuard}
\label{method}
FaceSwapGuard (\texttt{FSG}) protects against DeepFake face-swapping by adding imperceptible perturbations to user photos, creating protected images. When adversaries use these images for face swapping, the generated images' identity features significantly deviate from the source, preventing identity theft. Additionally, the visual differences between generated and source images reduce privacy risks from fake information dissemination. In summary, \texttt{FSG}'s protected images effectively mitigate face-swapping threats.

\begin{figure*}[t!]
\centering
\includegraphics[width=0.85\linewidth]{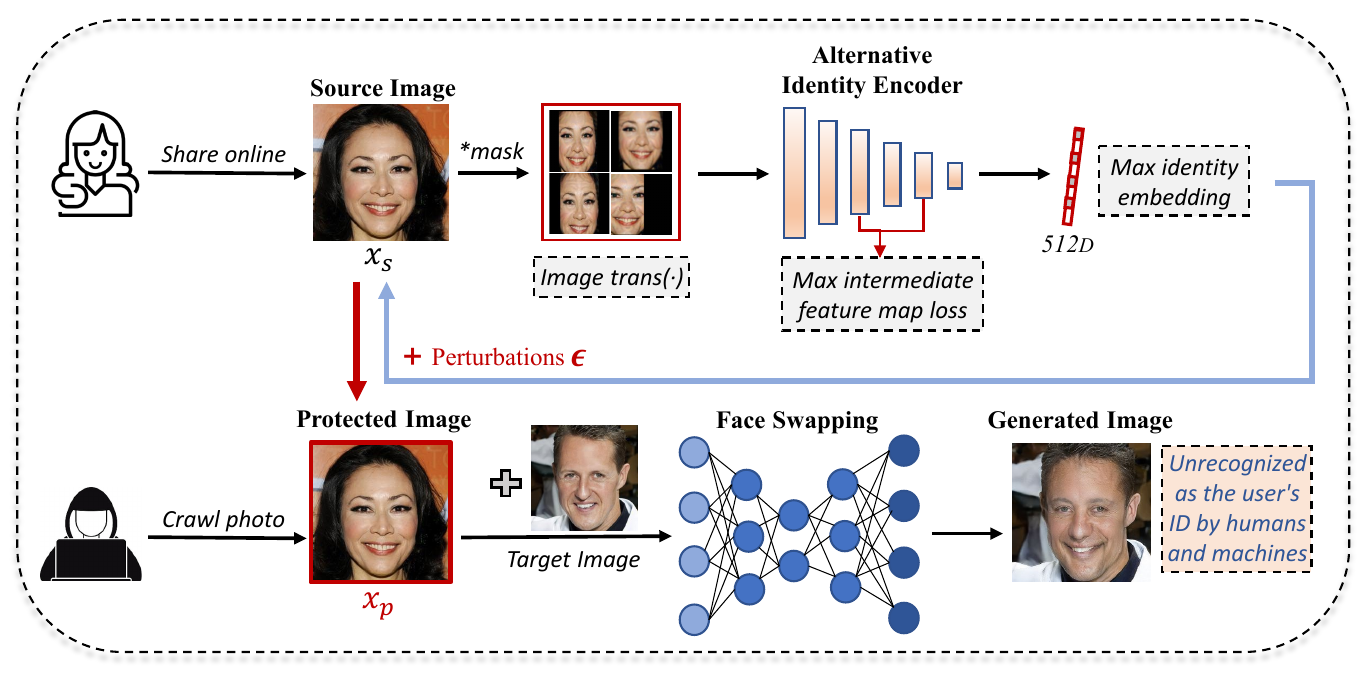} 
\caption{The overview framework of our \texttt{FSG}. Before sharing photos on social media platforms, users can utilize \texttt{FSG} to compute corresponding protected image by iteratively adding
adversarial perturbations to their source image. When an adversary scrapes the protected image for malicious DeepFake, the resulting face-swapped image deviates from the user's identity, making it difficult for both machines and humans to accurately match
the user’s genuine identity. }
\label{overview2}
\end{figure*}

\subsection{Motivation and Intuition}
\label{overview and intuition}
Our motivation is twofold: First, DeepFake threats, especially face swapping, pose serious risks to identity authentication systems~\cite{li2022seeing}.  
Second, existing pixel-level defenses fail to mitigate these identity thefts caused by face-swapping, as they only degrade image quality without altering semantic features, especially in black-box scenarios.  
Further analyses are provided in Appendix \ref{A1}.

To counter DeepFake face swapping, we propose FaceSwapGuard (\texttt{FSG}), a proactive defense based on identity obfuscation.  
Our approach disrupts feature extraction by maximizing deviations in identity encoder outputs, preventing the swapped face from matching the source.  
\autoref{overview2} provides an overview of \texttt{FSG}.

\subsection{Protected Image Computation}
\label{Computing protected image}
Individual users can apply \texttt{FSG} to generate protected images that closely resemble the source image while disrupting identity extraction during face swapping. The design goals are:  
\begin{itemize}
    \item The protected image should be visually identical to the source image. 
    \item Face-swapped images generated from the protected image should have identity features and visual appearances significantly different from the source. 
\end{itemize}
Generally, \texttt{FSG} aims to maximize identity deviation in feature space while preserving human perceptual differences. A key challenge is ensuring protection remains effective across various unknown face-swapping models. To address this, \texttt{FSG} is designed for transferability and robustness. The following sections detail the methodology for generating protected images.

\subsubsection{Maximizing Identity features} 
Given the source image \(X_s\) that a user intends to share online, we iteratively search for a protected image \(X_p\) in the image space by modifying \(X_s\). The objective is to maximize the identity features deviation between \(X_s\) and \(X_p\) as follows:

\begin{equation}
\begin{gathered}
\text{max}_{x_p}D\left (E\left (x_{p} \odot M \right ) ,E\left (x_{s} \odot M \right ) \right )  \\
\text{s.t.} \left|x_{p}-x_{s} \right|_{\infty}< \epsilon 
\label{equ1}
\end{gathered}
\end{equation}
where \(E(\cdot )\) represents the identity features extracted from the identity encoder, \(\left|x_{p}-x_{s} \right|_{\infty}\) measures the distance between \(X_p\) and \(X_s\), \(\epsilon\) is the distance budget in image space,  \(M\) denotes the facial region mask, 
and \(D(\cdot )= (1-cos(\cdot,\cdot ))+mse(\cdot,\cdot)\) computes the distance of two feature vectors. Here, \(cos(\cdot ,\cdot )\) is the cosine similarity, and \(mse(\cdot ,\cdot)\) is the MSE loss. 

Besides, we introduce intermediate feature map loss ~\cite{wang2018high} to enhance identity feature distortion during face swapping, increasing the visual difference between generated and source images. 
Inspired by SimSwap ~\cite{chen2020simswap}, we use feature maps from the last few layers, which capture high-level identity semantics, to compute the loss. 
The formula is as follows:
\begin{equation}
\begin{gathered}
\text{max}_{x_{p}} \sum_{i=k}^{K} D\left (E_i\left ( x_{p} \odot M \right )  , E_i\left (x_{s} \odot M \right ) \right ) \\
\text{s.t.} \left|x_{p}-x_{s} \right|_{\infty}< \epsilon 
\label{equ2}
\end{gathered}
\end{equation}
where \(K\) is the total number of layers in the identity encoder, and \(k\) is the first layer where we calculate feature map loss.

\subsubsection{Transferability and Robustness} 
If the defender, i.e., the \texttt{FSG} user, knows the adversary’s identity encoder, Equation \ref{equ2} can be used directly to distort identity features. However, in a general black-box scenario where the defender lacks access to the adversary’s encoder or the face-swapping model lacks explicit identity constraints, we construct a surrogate identity encoder. This improves the transferability and robustness of protected images across different face-swapping models.

The identity encoder in a face-swapping model extracts identity features from \(X_s\) and preserves them during generation to ensure consistency. Thus, we use open-source face recognition models, denoted as \(F\), as the surrogate encoder.

To enhance the transferability and robustness of protected images, we employ a random image transformation function during the optimization process ~\cite{yang2021defending}. These transformations simulate common operations encountered in face swapping, such as resizing and cropping. The whole process remains differentiable, and we reformulate the optimization function as follows:
\begin{equation}
\begin{gathered}
\text{max}_{x_{p}} \sum_{i=k}^{K} D\left ( F_i\left ( Trans\left (x_{p}\right ) \odot M \right )  , F_i\left (x_{s} \odot M \right ) \right ) \\
\text{s.t.} \left|x_{p}-x_{s} \right|_{\infty}< \epsilon 
\label{equ3}
\end{gathered}
\end{equation}
where \(F\left (\cdot \right )\) is the surrogate model and \(Trans\left (\cdot \right )\) is the random transformation function, detailed in Appendix \ref{A2}.

Algorithm \ref{alg1} presents the overall pseudocode of \texttt{FSG} based on the projected gradient descent (PGD) ~\cite{madry2017pgd}. Throughout the iterative process, we apply small-scale random padding within facial region \( M\) to adapt to various facial detection pre-processing. \(Clip_ \epsilon\) and \(Clip_{image}\) represent the clipping in the $\epsilon$-ball and the image range. 

\subsection{Face-swapped Image Generation}
\label{threat model}

\subsubsection{Threat model}
The adversary aims to perform unauthorized face swapping on users' photos for malicious purposes, such as bypassing face verification systems to access secure applications or spreading false information that invades privacy. Adversaries can range from commercial entities to malicious individuals, and we assume they have access to advanced face-swapping models \(G(\cdot)\) to manipulate any user's photos.

\subsubsection{Face swapping under protection} 
In practical applications, users can apply Algorithm \ref{alg1} to generate a protected image by iteratively adding adversarial perturbations to their source image before sharing it on social media. 
This ensures that any face-swapped image will have significant differences in identity features, making it impossible for both machines and humans to recognize the user's true identity, thereby thwarting adversarial manipulation. 
The formula is as follows:
\begin{equation}
\begin{gathered}
F\left ( G\left ( x_{p} \right )  \right ) \ne F\left ( G\left ( x_{s} \right )  \right ) 
\label{equ4}
\end{gathered}
\end{equation}

\begin{algorithm}[tb]
\caption{Search the protected images using our \texttt{FSG}.}
\label{alg1}
\textbf{Input}: source image $x_{s}$, pre-trained surrogate model \(F\left (\cdot \right )\),  random transformation function \(Tran\left (\cdot \right )\),  facial region mask \(M\), iterations $Iter$, perturbation budget \(\epsilon\), step size \(\alpha\).\\
\textbf{Output}:  the protected image \(x_{p}\)
\begin{algorithmic}[1] 
\State Initialize $x_{p}$ with random perturbation under budget $\epsilon$.
\State Initialize \(Trans\left (\cdot \right )\) with random blurring, resizing and cropping.
\For{ $j=1,...,Iter$}
\State \(M_j=Pad_{random}(M)\)
\State \small $\begin{aligned}
Loss=\sum\limits_{i=k}^{K} D\left ( F_i\left ( Trans\left (x_{p}\right ) \odot M_j\right )  , F_i\left (x_{s} \odot M_j \right ) \right )
\end{aligned}$
\State $noise=Clip_\epsilon \left ( x_{p}-x_{s}+\alpha \cdot sign\left (\bigtriangledown_{x_p} Loss\right )  \right ) $
\State $x_{p}=Clip_{image} \left ( x_{s} + noise\right ) $
\EndFor
\State \textbf{return}  $x_{p}$
\end{algorithmic}
\end{algorithm}

\section{Evaluation}
\label{evaluation}

\subsection{Experimental Setup}
\textbf{Datasets.} Our goal is to protect users' source images from face-swapping threats. Face swapping typically requires a source image for identity information and a target image for attributes and background. In our experiments, we randomly select 100 source images with different identities from the CelebA-HQ~\cite{karras2017celeb-hq} dataset as potential victims. We also randomly choose 10 target images from the remaining identities, allowing each face-swapping model to generate 1,000 face-swapped images.

\textbf{Models.} (1) Face-swapping Models: We select two feature-level face-swapping methods, FaceShifter and SimSwap, both of which use identity encoders to extract features from source images. These models have shown strong performance in bypassing facial liveness verification ~\cite{li2022seeing}. Additionally, we assess the effectiveness of \texttt{FSG} on other DeepFake models, including diffusion-based models, as detailed in Appendix \ref{A6}.
(2) Surrogate Models for Identity Encoder: We use the open-source face recognition models FaceNet ~\cite{schroff2015facenet} and ArcFace ~\cite{deng2019arcface} as surrogate models, representing different training loss functions ~\cite{yang2020robfr}. FaceNet is trained with an Euclidean-distance-based loss on the 112×112 aligned MS-Celeb-1M dataset ~\cite{guo2016ms}, while ArcFace employs angular-margin-based losses and is trained on the VGGFace2 dataset ~\cite{cao2018vggface2}.

\textbf{Metrics.}
(1) Face Matching Rate (FMR): A successful defense is indicated by identity deviation in face-swapped images compared to the source image. We evaluate our method using three academic face recognition models (FaceNet, ArcFace, and CosFace~\cite{wang2018cosface}) and three industrial face verification APIs (Baidu \footnote{https://cloud.baidu.com/product/face/compare.}, Tencent \footnote{https://cloud.tencent.com/product/facerecognition.}, and Face++ \footnote{https://www.faceplusplus.com.cn/face-comparing/.}).
If the feature distance between the generated image \(x_g\) and the source image exceeds a threshold, the identity is deemed different, indicating a successful defense. We evaluate this using the average FMR across \(N=1,000\) face pairs:
\begin{equation}
\begin{aligned}
FMR\left ( \% \right ) = 100\times \frac{1}{N}\sum_{i=1}^{N}\left [ \left\| F\left ( x_{g}^{i} \right )-F\left ( x_{s}^{i} \right )\right\| < T_{f}\right ]
\label{equ5}
\end{aligned}
\end{equation}
Therefore, a lower matching rate indicates a higher effectiveness of our method, which means more generated images deviate from the identity of the source images.

(2) Perceptual Similarity (LPIPS): To quantitatively assess the visual difference between the generated and source images, we calculate the average LPIPS~\cite{zhang2018lpips} distance for 100 source faces and 1,000 generated faces. 
We also provide qualitative visualizations to demonstrate the human perceptual differences.

\subsection{Baseline Results}
We first demonstrate the identity theft threat posed by face swapping without protection. We generate 1,000 face-swapped images using FaceShifter~\cite{li2019FaceShifter} and SimSwap~\cite{chen2020simswap} to establish baseline results.
\begin{table}[!t]
\caption{FMRs between normally generated and source images.}
\centering
\scalebox{0.85}{
\setlength{\tabcolsep}{4pt}
\begin{tabular}{c|lll}
\hline 
\toprule[1pt]
\multicolumn{1}{l}{}  &  & FaceShifter & SimSwap \\ \hline 
\multirow{3}{*}{\begin{tabular}
[c]{@{}c@{}}Face recognition \\ Models\end{tabular}} 
& ArcFace  & \multicolumn{1}{c}{96.2\%}       & \multicolumn{1}{c}{93.9\%}    \\ 
& FaceNet & \multicolumn{1}{c}{99.7\%}        & \multicolumn{1}{c}{ 99.2\%}  \\ 
& CosFace & \multicolumn{1}{c}{87.2\%}      & \multicolumn{1}{c}{81.3\%}  \\ \hline \hline
\multirow{3}{*}{\begin{tabular}[c]{@{}c@{}}Face verification \\ APIs\end{tabular}}  
& Baidu   & \multicolumn{1}{c}{91.6\%}        & \multicolumn{1}{c}{94.4\%}    \\ 
& Tencent & \multicolumn{1}{c}{95.4\%}        & \multicolumn{1}{c}{97.9\%}   \\ 
& Face++  & \multicolumn{1}{c}{99.3\%}       & \multicolumn{1}{c}{99.9\%}   \\ \hline 
\toprule[1pt]
\end{tabular}}
\vskip -0.1in
\label{baseline1}
\end{table}
\autoref{baseline1} shows the FMRs of both normally generated and source images across various face recognition models and commercial APIs. The FMRs between face-swapped and source images exceed 90\% for all APIs, highlighting a significant identity theft threat. 
In \autoref{baseline2}, we present the visual perception results of the generated images, combining quantitative LPIPS distance with qualitative visualizations. Despite minor variations between models, the generated images retain identity features similar to the source image, making them susceptible to malicious use, such as spreading fake news and damaging reputations. 

\begin{table}[t]
    \caption{Quantitative and qualitative perceptual similarity of the normally generated images.}
    \centering
    \scalebox{0.85}
    {
    \begin{tabular}{c|c|c|c}
    \toprule[1pt]
       Source Image & Target Image &  FaceShifter &  SimSwap\\
         \midrule
        \raisebox{-.5\height}{\includegraphics[width=0.19\linewidth]{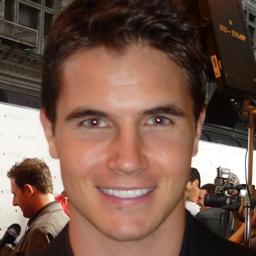}} & 
        \raisebox{-.5\height}{\includegraphics[width=0.19\linewidth]{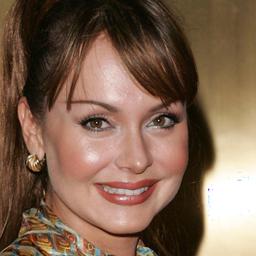}} & 
        \raisebox{-.5\height}{\includegraphics[width=0.19\linewidth]{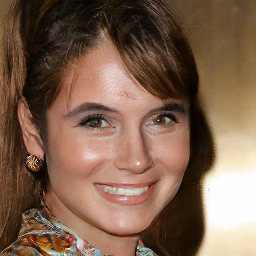}} &
        \raisebox{-.5\height}{\includegraphics[width=0.19\linewidth]{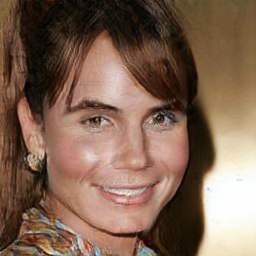}}
        \\
         \midrule
        \raisebox{-.5\height}{\includegraphics[width=0.19\linewidth]{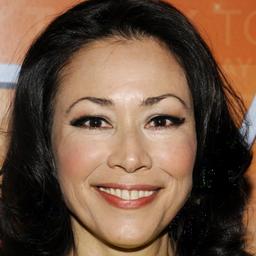}} & 
        \raisebox{-.5\height}{\includegraphics[width=0.19\linewidth]{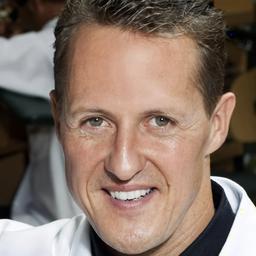}} & 
        \raisebox{-.5\height}{\includegraphics[width=0.19\linewidth]{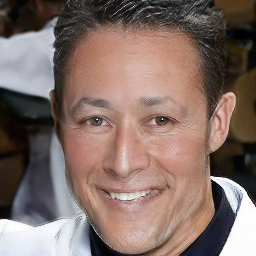}} &
        \raisebox{-.5\height}{\includegraphics[width=0.19\linewidth]{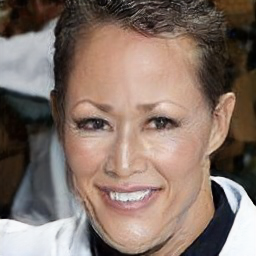}}
        \\
         \bottomrule
        \multicolumn{2}{c|} {LPIPS} &  0.5426 &  0.5368\\ 
        \toprule[1pt]
    \end{tabular}}
    \label{baseline2}
\end{table}

\subsection{Effectiveness Under Black-box Scenario }
We use \texttt{FSG} to generate protected images with different surrogate models and evaluate their effectiveness against various unseen face-swapping techniques.
\begin{table}[t]
\caption{FMRs between the generated images and the source images under the black-box scenario with \texttt{FSG} protection.}
\centering
\scalebox{0.9}{
\setlength{\tabcolsep}{4pt}
\begin{tabular}{clll}
\hline
\toprule[1pt]
\multicolumn{4}{c}{Surrogate model: ArcFace} \\ \hline
\multicolumn{1}{l}{}   & \multicolumn{1}{l}{}        
& \multicolumn{1}{l}{FaceShifter} & SimSwap \\ \hline
\multicolumn{1}{c|}{\multirow{3}{*}
{\begin{tabular}[c]{@{}c@{}}Face recognition \\ Models\end{tabular}}} 
& \multicolumn{1}{l}{ArcFace} 
& \multicolumn{1}{c}{1.9\%}    & \multicolumn{1}{c}{4.3\%}    \\ 
\multicolumn{1}{c|}{}  & \multicolumn{1}{l}{FaceNet} 
& \multicolumn{1}{c}{3.6\%}    & \multicolumn{1}{c}{ 6.5\%}  \\ 
\multicolumn{1}{c|}{}  & \multicolumn{1}{l}{CosFace} 
& \multicolumn{1}{c}{4.1\%}    & \multicolumn{1}{c}{ 8\% } \\ \hline \hline
\multicolumn{1}{c|}{\multirow{3}{*}
{\begin{tabular}[c]{@{}c@{}}Face verification \\ APIs\end{tabular}}}  
& \multicolumn{1}{c}{Baidu}   
& \multicolumn{1}{c}{0}        & \multicolumn{1}{c}{ 0}    \\ 
\multicolumn{1}{c|}{}  & \multicolumn{1}{c}{Tencent}
& \multicolumn{1}{c}{1.7\%}       & \multicolumn{1}{c}{ 5.2\%}   \\  
\multicolumn{1}{c|}{}  & \multicolumn{1}{c}{Face++}  
& \multicolumn{1}{c}{2.5\%}       & \multicolumn{1}{c}{6.7\% }  \\ \hline \toprule[1pt]

\multicolumn{4}{c}{Surrogate model: FaceNet}  \\ \hline
\multicolumn{1}{c|}{\multirow{3}{*}
{\begin{tabular}[c]{@{}c@{}}Face recognition \\ Models\end{tabular}}} 
& \multicolumn{1}{c}{ArcFace} 
& \multicolumn{1}{c}{10.6\%}     & \multicolumn{1}{c}{23.8\%}    \\ 
\multicolumn{1}{c|}{}  & \multicolumn{1}{c}{FaceNet} 
& \multicolumn{1}{c}{0.1\%}      & \multicolumn{1}{c}{ 2.3\%}  \\  
\multicolumn{1}{c|}{}  & \multicolumn{1}{c}{CosFace} 
& \multicolumn{1}{c}{8.1\%}      & \multicolumn{1}{c}{15.5\%}  \\ \hline \hline
\multicolumn{1}{c|}{\multirow{3}{*}
{\begin{tabular}[c]{@{}c@{}}Face verification \\ APIs\end{tabular}}}  
& \multicolumn{1}{c}{Baidu}   
& \multicolumn{1}{c}{0}        & \multicolumn{1}{c}{ 0.1\% }   \\  
\multicolumn{1}{c|}{}  & \multicolumn{1}{c}{Tencent}
& \multicolumn{1}{c}{2.1\%}    & \multicolumn{1}{c}{ 12.2\%}   \\  
\multicolumn{1}{c|}{}  & \multicolumn{1}{c}{Face++}  
& \multicolumn{1}{c}{3.6\%}    & \multicolumn{1}{c}{ 16.7\%}   \\ \hline
\toprule[1pt]
\end{tabular}}
\label{black1}
\end{table}

\subsubsection{Evaluation on Identity Obfuscation} 
We generate protected images using the surrogate models ArcFace and FaceNet, with a distance budget of \(\epsilon=9\) for pixel values in [0, 255]. 
These protected images are then used by FaceShifter and SimSwap to produce face-swapped images. 
\autoref{black1} shows the FMRs between the generated and source images across various face recognition models and verification APIs. 
Results demonstrate that \texttt{FSG} effectively mitigates face-swapping threats in black-box scenarios, significantly reducing FMRs. Key findings include:  
1) All FMRs are below 25\%, with real-world APIs dropping below 17\%.  
2) \texttt{FSG} performs best on Baidu’s API, reducing FMRs from over 90\% to nearly 0.  
3) Protected images based on ArcFace outperform those from FaceNet, with FMRs below 8\% across models and APIs, likely because ArcFace shares a backbone with the identity encoders used in the experiments.

\begin{figure}[!t]
\centering
\begin{subfigure}{0.24\textwidth}  
  \centering
  \includegraphics[width=\linewidth]{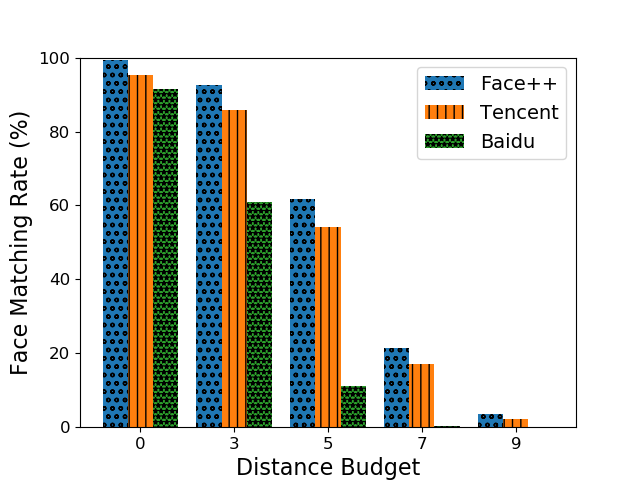}
  \caption{FaceShifter}
  \label{black-eps-fs}
\end{subfigure}
\hspace{-10pt}
\begin{subfigure}{0.24\textwidth}
  \centering
  \includegraphics[width=\linewidth]{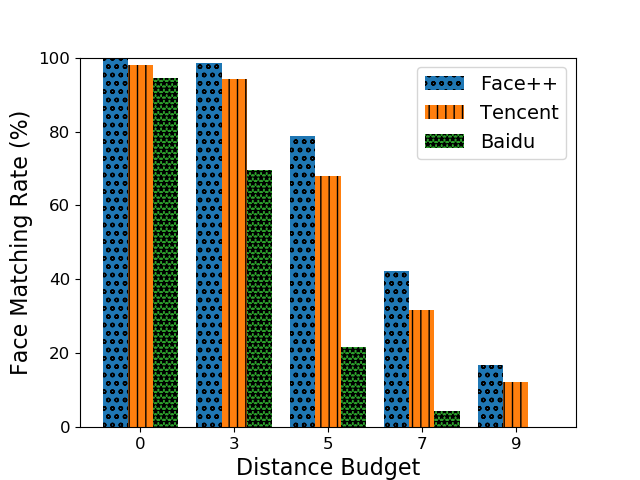}
  \caption{SimSwap}
  \label{black-eps-ss}
\end{subfigure}
\caption{FMRs on face verification APIs under different budgets.}
\label{black-eps}
\end{figure}

Furthermore, we evaluate the performance of protected images under different distance budgets \(\epsilon\). 
As shown in \autoref{black-eps}, we test a more realistic scenario where protected images are generated using the FaceNet surrogate model (which differs significantly from the identity encoders in the experimental face-swapping models) and calculate the FMRs on commercial face verification APIs. 
Additional results using different face recognition models are provided in Appendix \ref{A3}. 
The FMRs of the face-swapped images decrease as the perturbation budgets increase, suggesting that the perturbations introduced by \texttt{FSG} effectively mitigate the threat posed by face-swapping to identity verification systems relying on facial recognition technology.

\begin{table}[!t]
\caption{Visualizations of protected images at different budgets.}
    \centering
    \setlength{\tabcolsep}{1pt}
    \scalebox{0.85}
    {
    \begin{tabular}{c|c|c|c|c}
    \toprule[1pt]
       Source Image & $\epsilon=3$ &  $\epsilon=5$ &  $\epsilon=7$  & $\epsilon=9$\\
         \midrule
        \raisebox{-.5\height}{\includegraphics[width=0.19\linewidth]{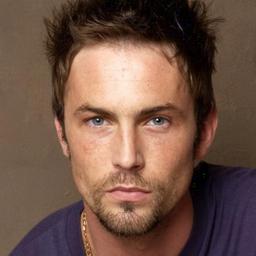}} & 
        \raisebox{-.5\height}{\includegraphics[width=0.19\linewidth]{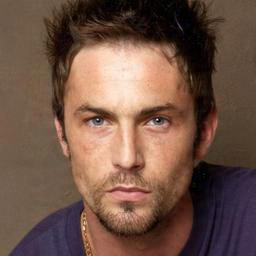}} & 
        \raisebox{-.5\height}{\includegraphics[width=0.19\linewidth]{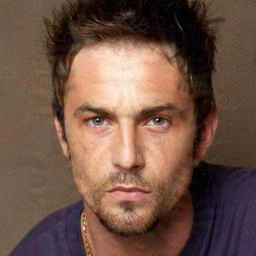}} &
        \raisebox{-.5\height}{\includegraphics[width=0.19\linewidth]{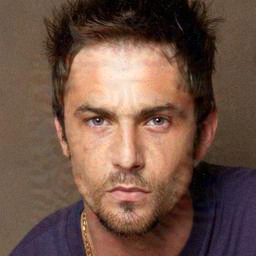}} &
        \raisebox{-.5\height}{\includegraphics[width=0.19\linewidth]{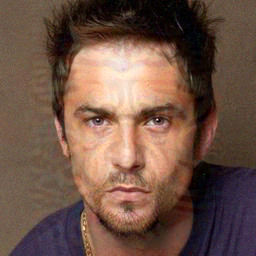}}
         \\

          \midrule
        \raisebox{-.5\height}{\includegraphics[width=0.19\linewidth]{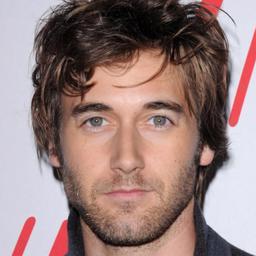}} & 
        \raisebox{-.5\height}{\includegraphics[width=0.19\linewidth]{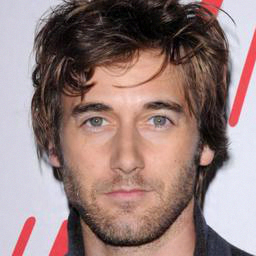}} & 
        \raisebox{-.5\height}{\includegraphics[width=0.19\linewidth]{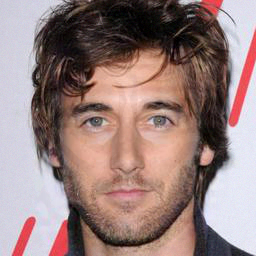}} &
        \raisebox{-.5\height}{\includegraphics[width=0.19\linewidth]{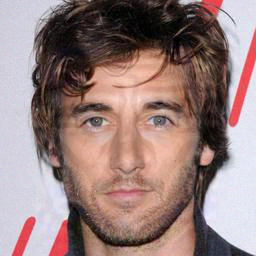}} &
        \raisebox{-.5\height}{\includegraphics[width=0.19\linewidth]{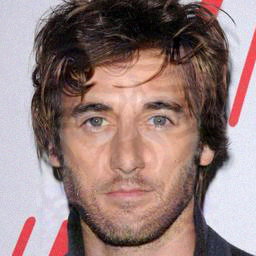}}
        \\
        \toprule[1pt]
    \end{tabular}}
    \label{black-adv}
\end{table}

Additionally, we present the visualization results of the protected images under different budgets in \autoref{black-adv}. When the budget is below 7, the perturbation remains imperceptible, and \texttt{FSG} achieves satisfactory effectiveness. 
Although some perturbations may become visually detectable at \(\epsilon = 9\), our approach still offers robust defense with a small perturbation value, compared to commonly used settings in black-box adversarial attacks (e.g., \(\epsilon = 10/15/16\))~\cite{zhong2020towards}. 

\subsubsection{Evaluation on Visual Confusion} 
In \autoref{black-gen-arcface}, we present the perceptual results at \(\epsilon = 9\) to demonstrate the effectiveness of \texttt{FSG} in altering visual perception.
Compared to the baseline results in \autoref{baseline2}, the LPIPS losses are higher, indicating larger feature differences. 
We also include visualization examples highlighting the differences between face-swapped images and the source images (additional results are in Appendix \ref{A4}). 
For an intuitive comparison, \autoref{black-gen-arcface} displays face-swapped images generated with the same target images as those in \autoref{baseline2}. However, our method is target-agnostic and maintains robust defense when applied to arbitrary target images. 
These results demonstrate that \texttt{FSG} creates images with noticeable feature differences, effectively mitigating misinformation and protecting individual users' identities.

\begin{table}[t]
  \caption{Quantitative and qualitative perceptual similarity of face-swapped images.}
    \centering
    \scalebox{0.85}
    {
    \begin{tabular}{c|c|c|c}
    \toprule[1pt]
       Source Image & Target Image &  FaceShifter &  SimSwap\\
         \midrule
        \raisebox{-.5\height}{\includegraphics[width=0.2\linewidth]{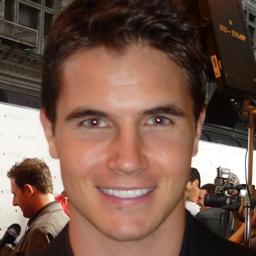}} & 
        \raisebox{-.5\height}{\includegraphics[width=0.2\linewidth]{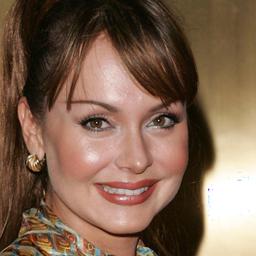}} & 
        \raisebox{-.5\height}{\includegraphics[width=0.2\linewidth]{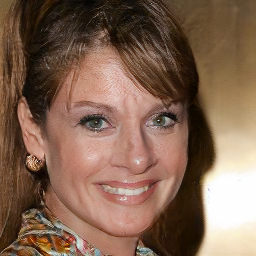}} &
        \raisebox{-.5\height}{\includegraphics[width=0.2\linewidth]{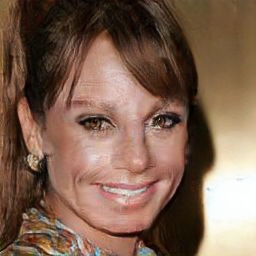}}
        \\
         \midrule
        \raisebox{-.5\height}{\includegraphics[width=0.2\linewidth]{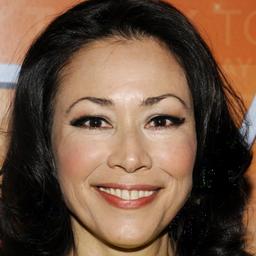}} & 
        \raisebox{-.5\height}{\includegraphics[width=0.2\linewidth]{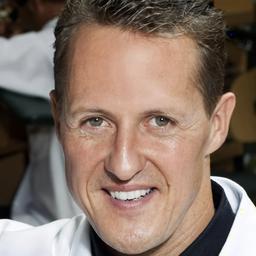}} & 
        \raisebox{-.5\height}{\includegraphics[width=0.2\linewidth]{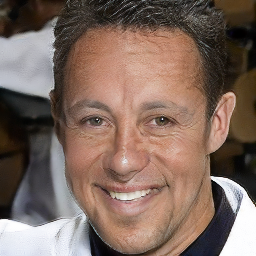}} &
        \raisebox{-.5\height}{\includegraphics[width=0.2\linewidth]{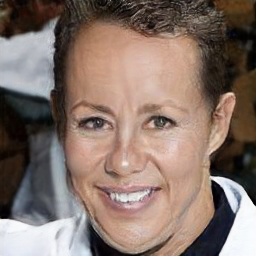}}
        \\
         \bottomrule
        \multicolumn{2}{c|} {LPIPS} &  0.5486 &  0.5397\\ 
        \toprule[1pt]
    \end{tabular}}
    \label{black-gen-arcface}
\end{table}

\subsection{Comparison with Pixel-level Disruption Method}
To further assess the effectiveness of \texttt{FSG}, we compare it with pixel-level disruption methods. Since Ruiz et al.'s method~\cite{ruiz2020disrupting} is designed for white-box scenarios and the lack of open-source code for black-box methods based on face-swapping models~\cite{dong2023restricted}, we evaluate existing defenses by maximizing the distance between adversarial and original images, as outlined in~\cite{ruiz2020disrupting, yeh2020disrupting}. 
We generate protected images using various face-swapping models and evaluate the transferability of adversarial examples to test the pixel-level disruption method across different models. All other experimental settings, such as image transformation functions and input data, remain consistent with \texttt{FSG}.

\begin{figure}[tt]
\centering
\begin{subfigure}{0.24\textwidth} 
  \centering
  \includegraphics[width=\linewidth]{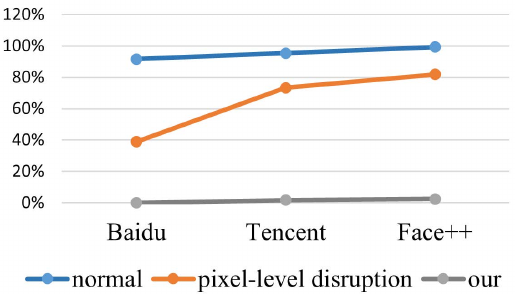}
  \caption{FaceShifter}
  \label{faceshifter_campare}
\end{subfigure}
\hspace{-10pt} 
\begin{subfigure}{0.24\textwidth}
  \centering
  \includegraphics[width=\linewidth]{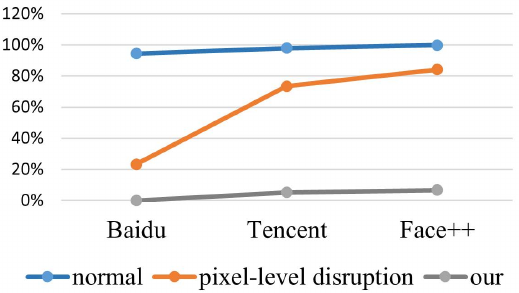}
  \caption{SimSwap}
  \label{simswap_campare}
\end{subfigure}
\caption{FMRs between the generated images and source images under different defenses.}
\label{compare_FMR}
\end{figure}

\begin{figure*}[!t]
\centering
\includegraphics[width=1\linewidth]{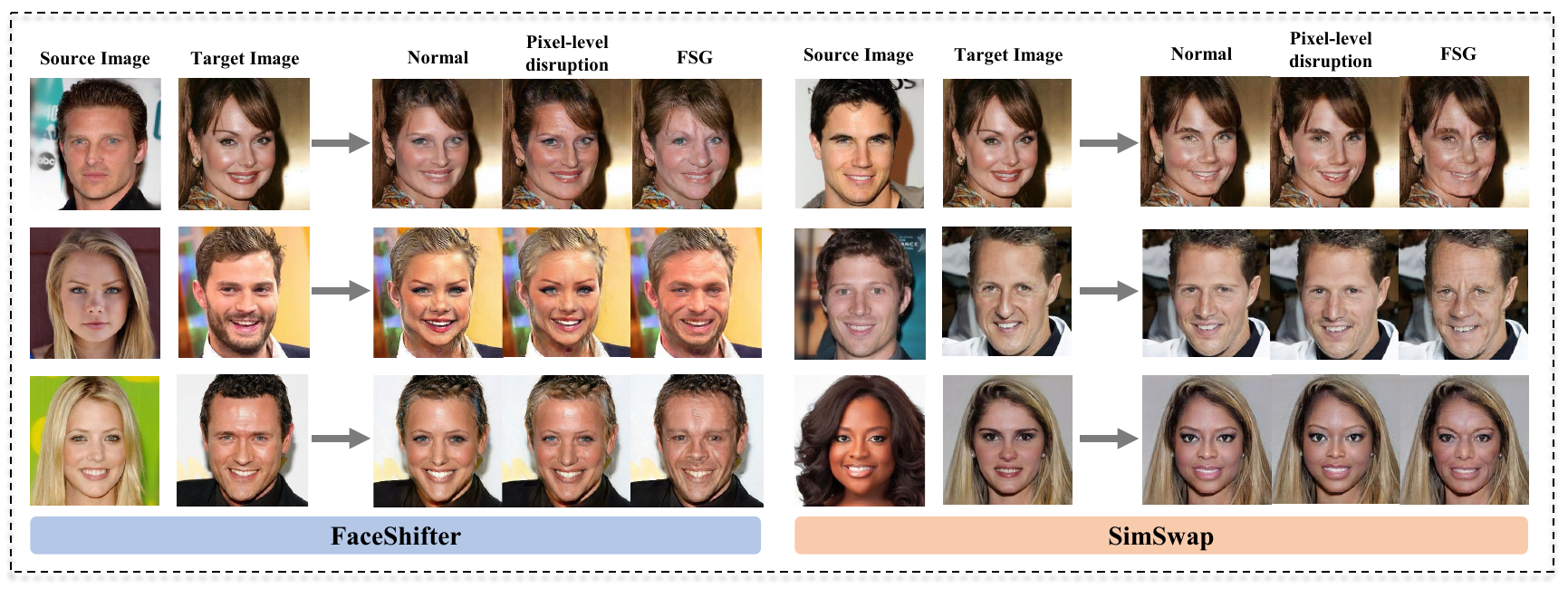} 
\caption{Visualization of face-swapped images under different defenses. }
\label{compare_visual}
\end{figure*}

As shown in \autoref{compare_FMR}, FMRs from various face verification APIs demonstrate that \texttt{FSG} achieves superior defense efficacy, with significantly lower FMRs. While pixel-level disruption methods show some identity confusion, their effect is limited. For instance, on Tencent and Face++, FMRs for pixel-level methods remain above 70\% and 80\%, respectively. In contrast, \texttt{FSG} reduces FMRs to below 7\%. Visual comparisons in \autoref{compare_visual} further confirm that \texttt{FSG} induces more perceptible identity confusion, creating greater deviations between face-swapped and source images. Thus, compared to pixel-level methods, \texttt{FSG} more effectively mitigates identity theft and misinformation risks posed by DeepFake.

\subsection{Ablation Study}
We perform ablation experiments to evaluate the contribution of each component in \texttt{FSG}, including the image transformation function and intermediate feature map loss. Three sets of experiments are designed: 
1) Removing random image transformations (``\textit{no\_Trans}''); 
2) Removing the intermediate feature map loss (``\textit{no\_Int.layer}''); 
3) Removing both random image transformations and intermediate feature map loss (``\textit{no\_Trans.\& Int.layer}'').

\begin{table}[!t]
\caption{FMRs under various settings and face recognition models.}
\centering
\scalebox{0.95}{
\setlength{\tabcolsep}{3pt}
\begin{tabular}{c|c|c c}
\hline \toprule[1pt]
\multicolumn{1}{c}{}  &    \multicolumn{1}{c}{}  & FaceShifter $\downarrow$      & SimSwap $\downarrow$        \\ \hline
\multirow{4}{*}{ArcFace} & no\_Trans             & 28.5\%          & 49\%            \\ 
                         & no\_Int.layer         & 14.6\%          & 33.5\%          \\  
                         & no\_Trans\&Int.layer & 44\%            & 60.8\%          \\  
                         &\texttt{FSG}           & \textbf{10.6\%} & \textbf{23.8\%} \\ \hline \hline
\multirow{4}{*}{FaceNet} & no\_Trans             & 3.8\%           & 20.4\%          \\ 
                         & no\_Int.layer         & 0.1\%           & 4.5\%           \\ 
                         & no\_Trans\&Int.layer & 1\%             & 34.1\%          \\ 
                         &\texttt{FSG}           & \textbf{0.1\%}  & \textbf{2.3\%}  \\ \hline \hline
\multirow{4}{*}{CosFace} & no\_Trans             & 21.7\%          & 32.6\%          \\ 
                         & no\_Int.layer         & 10.7\%          & 22.6\%          \\ 
                         & no\_Trans\&Int.layer & 34.2\%          & 43.4\%          \\ 
                         & \texttt{FSG}          & \textbf{8.1\%}  & \textbf{15.5\%} \\ \hline \toprule[1pt]
\end{tabular}}

\label{ablation1}
\end{table}

We evaluate the effectiveness of protected images generated by the FaceNet surrogate model under various settings. 
As shown in \autoref{ablation1}, \texttt{FSG} achieves the lowest FMRs, effectively obfuscating identity and mitigating DeepFake threats to user privacy. 
Removing any component from \texttt{FSG} reduces its effectiveness, emphasizing the critical roles of random image transformations and intermediate feature map loss. 
These components enhance the transferability and robustness of protected images across different face-swapping models. 
Additionally, the perceptual similarity results in Table \ref{ablation2} demonstrate that \texttt{FSG} maximizes the visual perceptual difference between face-swapped images and the original source images.

\begin{table}[!t]
\caption{LPIPS distance under different settings.}
\centering
\scalebox{0.9}{
\setlength{\tabcolsep}{4pt}
\begin{tabular}{c|c c}
\hline \toprule[1.2pt]
\multicolumn{1}{c}{}  & FaceShifter $\uparrow$      & SimSwap $\uparrow$   \\ \hline  
no\_Trans            & 0.5440          & 0.5386         \\ 
no\_Int.layer        & 0.5457          & 0.5390         \\ 
no\_Trans\&Int.layer & 0.5442          & 0.5482         \\ 
\texttt{FSG}         & 0.5458          & 0.5397 \\ \hline 
\toprule[1.2pt]
\end{tabular}}
\label{ablation2}
\end{table}

\subsection{Adaptive Adversary}
\texttt{FSG} generates protected images by adding imperceptible adversarial noise to the source images. To test its robustness, we evaluate its performance against common image denoising techniques, including Gaussian blur and image compression ~\cite{kurakin2016adversarial}, which adversaries may use to purify the protected images.

\begin{figure}[!t]
\centering
\begin{subfigure}{0.24\textwidth}  
  \centering
  \includegraphics[width=\linewidth]{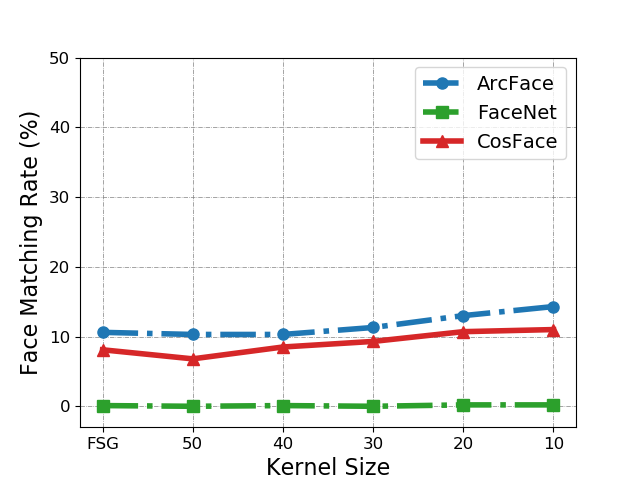}
  \caption{FaceShifter}
  \label{robust-blur-fs}
\end{subfigure}
\hspace{-10pt}  
\begin{subfigure}{0.24\textwidth}
  \centering
  \includegraphics[width=\linewidth]{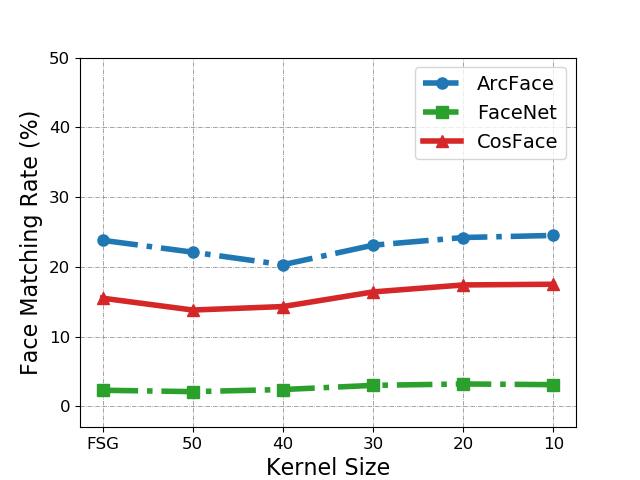}
  \caption{SimSwap}
  \label{robust-blur-ss}
\end{subfigure}
\caption{Face matching rates under Gaussian blur.}
\label{robust-blur}
\end{figure}

\begin{figure}[htbp]
\centering
\begin{subfigure}{0.24\textwidth} 
  \centering
  \includegraphics[width=\linewidth]{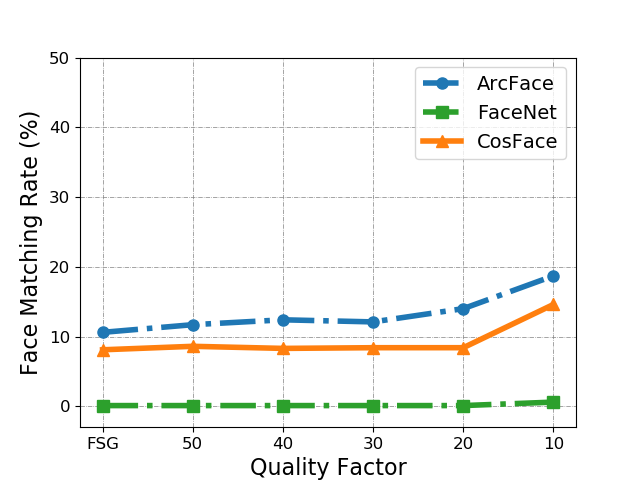}
  \caption{FaceShifter}
  \label{robust-qf-fs}
\end{subfigure}
\hspace{-10pt} 
\begin{subfigure}{0.24\textwidth}
  \centering
  \includegraphics[width=\linewidth]{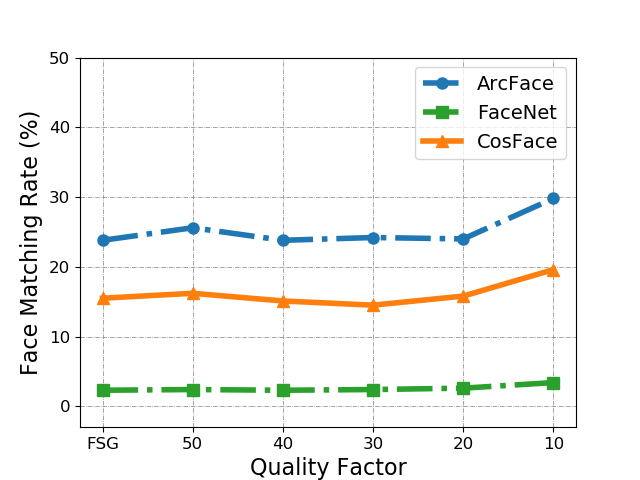}
  \caption{SimSwap}
  \label{robust-qf-ss}
\end{subfigure}
\caption{Face matching rates under image compression.}
\label{robust-qf}
\end{figure}

We apply Gaussian blurring and compression to the protected images and then use various face-swapping models to generate images based on these protected versions. \autoref{robust-blur} shows the FMRs between the face-swapped and source images after applying Gaussian blur with different kernel sizes. \autoref{robust-qf} presents the FMRs of protected images under compression with varying quality factors (QF), where a lower QF indicates poorer image quality. The results demonstrate that the FMRs remain stable and low under both operations, indicating that the protected images are robust against adaptive adversaries using image purification techniques.

\section{Conclusion}
We propose \texttt{FSG}, an efficient defense method against identity theft threats posed by DeepFake face-swapping.
Specifically, \texttt{FSG} protects an individual's identity by introducing imperceptible perturbations to user photos. 
In this way, \texttt{FSG} can effectively obfuscate both the identity and visual perception of face-swapped images, reducing the risks of identity theft and misinformation. 
Extensive experiments show \texttt{FSG}'s effectiveness in black-box scenarios, particularly in protecting against face-swapping attacks targeting commercial face-recognition APIs. 
We also verify the robustness of protected images against potential adaptive attacks.
Additionally, its success with both FSGAN and diffusion-based models demonstrates its potential for broader application across various DeepFake models. 
Overall, \texttt{FSG} offers a novel approach to safeguarding individual identity in the face of evolving DeepFake threats.

\section*{Impact Statements}
In this study, we perform a comprehensive security evaluation of face verification models and APIs against DeepFake face-swapping threats. In line with prior research on the security of AI-powered systems, we have carefully considered the legal and ethical implications of our work. First, we utilize open-source datasets for DeepFake synthesis and security assessment, a widely accepted and legitimate approach in face-related security research. Second, our evaluation of commercial face verification APIs adheres strictly to the official guidelines, with all API usage properly licensed and paid for. Additionally, we ensure that the number of queries per second (QPS) remains within the recommended limits to avoid disrupting the normal operation of the target service. Thus, our work does not raise ethical issues or negatively impact the target vendor's services.
\bibliographystyle{plain}
\bibliography{normal_generated}

\begin{thebibliography}{10}

\bibitem{bao2018IPGAN}
Jianmin Bao, Dong Chen, Fang Wen, Houqiang Li, and Gang Hua.
\newblock Towards open-set identity preserving face synthesis.
\newblock In {\em Proceedings of the IEEE conference on computer vision and pattern recognition}, pages 6713--6722, 2018.

\bibitem{cao2018vggface2}
Qiong Cao, Li~Shen, Weidi Xie, Omkar~M Parkhi, and Andrew Zisserman.
\newblock Vggface2: A dataset for recognising faces across pose and age.
\newblock In {\em 2018 13th IEEE international conference on automatic face \& gesture recognition (FG 2018)}, pages 67--74. IEEE, 2018.

\bibitem{chen2020simswap}
Renwang Chen, Xuanhong Chen, Bingbing Ni, and Yanhao Ge.
\newblock Simswap: An efficient framework for high fidelity face swapping.
\newblock In {\em Proceedings of the 28th ACM International Conference on Multimedia}, pages 2003--2011, 2020.

\bibitem{choi2018stargan}
Yunjey Choi, Minje Choi, Munyoung Kim, Jung-Woo Ha, Sunghun Kim, and Jaegul Choo.
\newblock Stargan: Unified generative adversarial networks for multi-domain image-to-image translation.
\newblock In {\em Proceedings of the IEEE conference on computer vision and pattern recognition}, pages 8789--8797, 2018.

\bibitem{deng2019arcface}
Jiankang Deng, Jia Guo, Niannan Xue, and Stefanos Zafeiriou.
\newblock Arcface: Additive angular margin loss for deep face recognition.
\newblock In {\em Proceedings of the IEEE/CVF conference on computer vision and pattern recognition}, pages 4690--4699, 2019.

\bibitem{dong2023restricted}
Junhao Dong, Yuan Wang, Jianhuang Lai, and Xiaohua Xie.
\newblock Restricted black-box adversarial attack against deepfake face swapping.
\newblock {\em IEEE Transactions on Information Forensics and Security}, 2023.

\bibitem{gu2022delving}
Zhihao Gu, Yang Chen, Taiping Yao, Shouhong Ding, Jilin Li, and Lizhuang Ma.
\newblock Delving into the local: Dynamic inconsistency learning for deepfake video detection.
\newblock In {\em Proceedings of the AAAI Conference on Artificial Intelligence}, pages 744--752, 2022.

\bibitem{guo2016ms}
Yandong Guo, Lei Zhang, Yuxiao Hu, Xiaodong He, and Jianfeng Gao.
\newblock Ms-celeb-1m: A dataset and benchmark for large-scale face recognition.
\newblock In {\em Computer Vision--ECCV 2016: 14th European Conference, Amsterdam, The Netherlands, October 11-14, 2016, Proceedings, Part III 14}, pages 87--102. Springer, 2016.

\bibitem{huang2021initiative}
Qidong Huang, Jie Zhang, Wenbo Zhou, Weiming Zhang, and Nenghai Yu.
\newblock Initiative defense against facial manipulation.
\newblock In {\em Proceedings of the AAAI Conference on Artificial Intelligence}, volume~35, pages 1619--1627, 2021.

\bibitem{jovanovic2022generative}
Mladan Jovanovic and Mark Campbell.
\newblock Generative artificial intelligence: Trends and prospects.
\newblock {\em Computer}, 55(10):107--112, 2022.

\bibitem{karras2017celeb-hq}
Tero Karras, Timo Aila, Samuli Laine, and Jaakko Lehtinen.
\newblock Progressive growing of gans for improved quality, stability, and variation.
\newblock {\em arXiv preprint arXiv:1710.10196}, 2017.

\bibitem{kim2022diffusionclip}
Gwanghyun Kim, Taesung Kwon, and Jong~Chul Ye.
\newblock Diffusionclip: Text-guided diffusion models for robust image manipulation.
\newblock In {\em Proceedings of the IEEE/CVF Conference on Computer Vision and Pattern Recognition}, pages 2426--2435, 2022.

\bibitem{korshunov2018deepfakes}
Pavel Korshunov and S{\'e}bastien Marcel.
\newblock Deepfakes: a new threat to face recognition? assessment and detection.
\newblock {\em arXiv preprint arXiv:1812.08685}, 2018.

\bibitem{kurakin2016adversarial}
Alexey Kurakin, Ian Goodfellow, Samy Bengio, et~al.
\newblock Adversarial examples in the physical world, 2016.

\bibitem{li2022seeing}
Changjiang Li, Li~Wang, Shouling Ji, Xuhong Zhang, Zhaohan Xi, Shanqing Guo, and Ting Wang.
\newblock Seeing is living? rethinking the security of facial liveness verification in the deepfake era.
\newblock In {\em 31st USENIX Security Symposium (USENIX Security 22)}, pages 2673--2690, 2022.

\bibitem{li2019FaceShifter}
Lingzhi Li, Jianmin Bao, Hao Yang, Dong Chen, and Fang Wen.
\newblock Faceshifter: Towards high fidelity and occlusion aware face swapping.
\newblock {\em arXiv preprint arXiv:1912.13457}, 2019.

\bibitem{madry2017pgd}
Aleksander Madry, Aleksandar Makelov, Ludwig Schmidt, Dimitris Tsipras, and Adrian Vladu.
\newblock Towards deep learning models resistant to adversarial attacks.
\newblock {\em arXiv preprint arXiv:1706.06083}, 2017.

\bibitem{masood2023trend1}
Momina Masood, Mariam Nawaz, Khalid~Mahmood Malik, Ali Javed, Aun Irtaza, and Hafiz Malik.
\newblock Deepfakes generation and detection: State-of-the-art, open challenges, countermeasures, and way forward.
\newblock {\em Applied intelligence}, 53(4):3974--4026, 2023.

\bibitem{meng2023ava}
Xiangtao Meng, Li~Wang, Shanqing Guo, Lei Ju, and Qingchuan Zhao.
\newblock Ava: Inconspicuous attribute variation-based adversarial attack bypassing deepfake detection.
\newblock {\em arXiv preprint arXiv:2312.08675}, 2023.

\bibitem{mirsky2021creation}
Yisroel Mirsky and Wenke Lee.
\newblock The creation and detection of deepfakes: A survey.
\newblock {\em ACM Computing Surveys (CSUR)}, 54(1):1--41, 2021.

\bibitem{nirkin2019fsgan}
Yuval Nirkin, Yosi Keller, and Tal Hassner.
\newblock Fsgan: Subject agnostic face swapping and reenactment.
\newblock In {\em Proceedings of the IEEE international conference on computer vision}, pages 7184--7193, 2019.

\bibitem{nirkin2018face}
Yuval Nirkin, Iacopo Masi, Anh~Tran Tuan, Tal Hassner, and Gerard Medioni.
\newblock On face segmentation, face swapping, and face perception.
\newblock In {\em 2018 13th IEEE International Conference on Automatic Face \& Gesture Recognition (FG 2018)}, pages 98--105. IEEE, 2018.

\bibitem{perez2003poisson}
Patrick P{\'e}rez, Michel Gangnet, and Andrew Blake.
\newblock Poisson image editing.
\newblock In {\em ACM SIGGRAPH 2003 Papers}, pages 313--318, 2003.

\bibitem{preechakul2022diffusion}
Konpat Preechakul, Nattanat Chatthee, Suttisak Wizadwongsa, and Supasorn Suwajanakorn.
\newblock Diffusion autoencoders: Toward a meaningful and decodable representation.
\newblock In {\em Proceedings of the IEEE/CVF Conference on Computer Vision and Pattern Recognition}, pages 10619--10629, 2022.

\bibitem{pumarola2018ganimation}
Albert Pumarola, Antonio Agudo, Aleix~M Martinez, Alberto Sanfeliu, and Francesc Moreno-Noguer.
\newblock Ganimation: Anatomically-aware facial animation from a single image.
\newblock In {\em Proceedings of the European conference on computer vision (ECCV)}, pages 818--833, 2018.

\bibitem{ruiz2020disrupting}
Nataniel Ruiz, Sarah~Adel Bargal, and Stan Sclaroff.
\newblock Disrupting deepfakes: Adversarial attacks against conditional image translation networks and facial manipulation systems.
\newblock In {\em Computer Vision--ECCV 2020 Workshops: Glasgow, UK, August 23--28, 2020, Proceedings, Part IV 16}, pages 236--251. Springer, 2020.

\bibitem{ruiz2020protecting}
Nataniel Ruiz, Sarah~Adel Bargal, and Stan Sclaroff.
\newblock Protecting against image translation deepfakes by leaking universal perturbations from black-box neural networks.
\newblock {\em arXiv preprint arXiv:2006.06493}, 2020.

\bibitem{schroff2015facenet}
Florian Schroff, Dmitry Kalenichenko, and James Philbin.
\newblock Facenet: A unified embedding for face recognition and clustering.
\newblock In {\em Proceedings of the IEEE conference on computer vision and pattern recognition}, pages 815--823, 2015.

\bibitem{tan2024rethinking}
Chuangchuang Tan, Yao Zhao, Shikui Wei, Guanghua Gu, Ping Liu, and Yunchao Wei.
\newblock Rethinking the up-sampling operations in cnn-based generative network for generalizable deepfake detection.
\newblock In {\em Proceedings of the IEEE/CVF Conference on Computer Vision and Pattern Recognition}, pages 28130--28139, 2024.

\bibitem{tariq2022real}
Shahroz Tariq, Sowon Jeon, and Simon~S Woo.
\newblock Am i a real or fake celebrity? evaluating face recognition and verification apis under deepfake impersonation attack.
\newblock In {\em Proceedings of the ACM Web Conference 2022}, pages 512--523, 2022.

\bibitem{tolosana2020deepfakes}
Ruben Tolosana, Ruben Vera-Rodriguez, Julian Fierrez, Aythami Morales, and Javier Ortega-Garcia.
\newblock Deepfakes and beyond: A survey of face manipulation and fake detection.
\newblock {\em Information Fusion}, 64:131--148, 2020.

\bibitem{verdoliva2020media}
Luisa Verdoliva.
\newblock Media forensics and deepfakes: an overview.
\newblock {\em IEEE Journal of Selected Topics in Signal Processing}, 14(5):910--932, 2020.

\bibitem{wang2018cosface}
Hao Wang, Yitong Wang, Zheng Zhou, Xing Ji, Dihong Gong, Jingchao Zhou, Zhifeng Li, and Wei Liu.
\newblock Cosface: Large margin cosine loss for deep face recognition.
\newblock In {\em Proceedings of the IEEE conference on computer vision and pattern recognition}, pages 5265--5274, 2018.

\bibitem{wang2024deepfaker}
Li~Wang, Xiangtao Meng, Dan Li, Xuhong Zhang, Shouling Ji, and Shanqing Guo.
\newblock Deepfaker: a unified evaluation platform for facial deepfake and detection models.
\newblock {\em ACM Transactions on Privacy and Security}, 27(1):1--34, 2024.

\bibitem{wang2024deepfake}
Tianyi Wang, Xin Liao, Kam~Pui Chow, Xiaodong Lin, and Yinglong Wang.
\newblock Deepfake detection: A comprehensive survey from the reliability perspective.
\newblock {\em ACM Computing Surveys}, 57(3):1--35, 2024.

\bibitem{wang2018high}
Ting-Chun Wang, Ming-Yu Liu, Jun-Yan Zhu, Andrew Tao, Jan Kautz, and Bryan Catanzaro.
\newblock High-resolution image synthesis and semantic manipulation with conditional gans.
\newblock In {\em Proceedings of the IEEE conference on computer vision and pattern recognition}, pages 8798--8807, 2018.

\bibitem{yang2021defending}
Chaofei Yang, Leah Ding, Yiran Chen, and Hai Li.
\newblock Defending against gan-based deepfake attacks via transformation-aware adversarial faces.
\newblock In {\em 2021 international joint conference on neural networks (IJCNN)}, pages 1--8. IEEE, 2021.

\bibitem{yang2020robfr}
Xiao Yang, Dingcheng Yang, Yinpeng Dong, Hang Su, Wenjian Yu, and Jun Zhu.
\newblock Robfr: Benchmarking adversarial robustness on face recognition.
\newblock {\em arXiv preprint arXiv:2007.04118}, 2020.

\bibitem{yeh2020disrupting}
Chin-Yuan Yeh, Hsi-Wen Chen, Shang-Lun Tsai, and Sheng-De Wang.
\newblock Disrupting image-translation-based deepfake algorithms with adversarial attacks.
\newblock In {\em Proceedings of the IEEE/CVF Winter Conference on Applications of Computer Vision Workshops}, pages 53--62, 2020.

\bibitem{zhang2018lpips}
Richard Zhang, Phillip Isola, Alexei~A Efros, Eli Shechtman, and Oliver Wang.
\newblock The unreasonable effectiveness of deep features as a perceptual metric.
\newblock In {\em Proceedings of the IEEE conference on computer vision and pattern recognition}, pages 586--595, 2018.

\bibitem{zhong2020towards}
Yaoyao Zhong and Weihong Deng.
\newblock Towards transferable adversarial attack against deep face recognition.
\newblock {\em IEEE Transactions on Information Forensics and Security}, 16:1452--1466, 2020.

\end{thebibliography}
\newpage
\appendix
\onecolumn
\section{Appendix}

\subsection{In-Depth Analysis of Existing Pixel-Level Disruption Methods} 
\label{A1}
Current pixel-level disruption methods show limited effectiveness in mitigating the threats posed by DeepFake attacks. These approaches primarily focus on degrading the visual quality of images at the pixel level. Especially in black-box scenarios, these methods often introduce only superficial artifacts without significantly altering the semantic features of the generated images.
To better illustrate the limitations of these methods, we conducted both quantitative and qualitative analyses. 
Since the disruption method proposed by Ruiz \textit{et al.} ~\cite{ruiz2020disrupting} is not suitable for face-swapping models in a black-box scenario, we follow their source code and conduct experiments for StarGAN ~\cite{choi2018stargan} and GANimation ~\cite{pumarola2018ganimation} in white-box scenarios, which similarly serves to illustrate the issue.

\begin{figure*}[ht]
\centering
\includegraphics[width=1\linewidth]{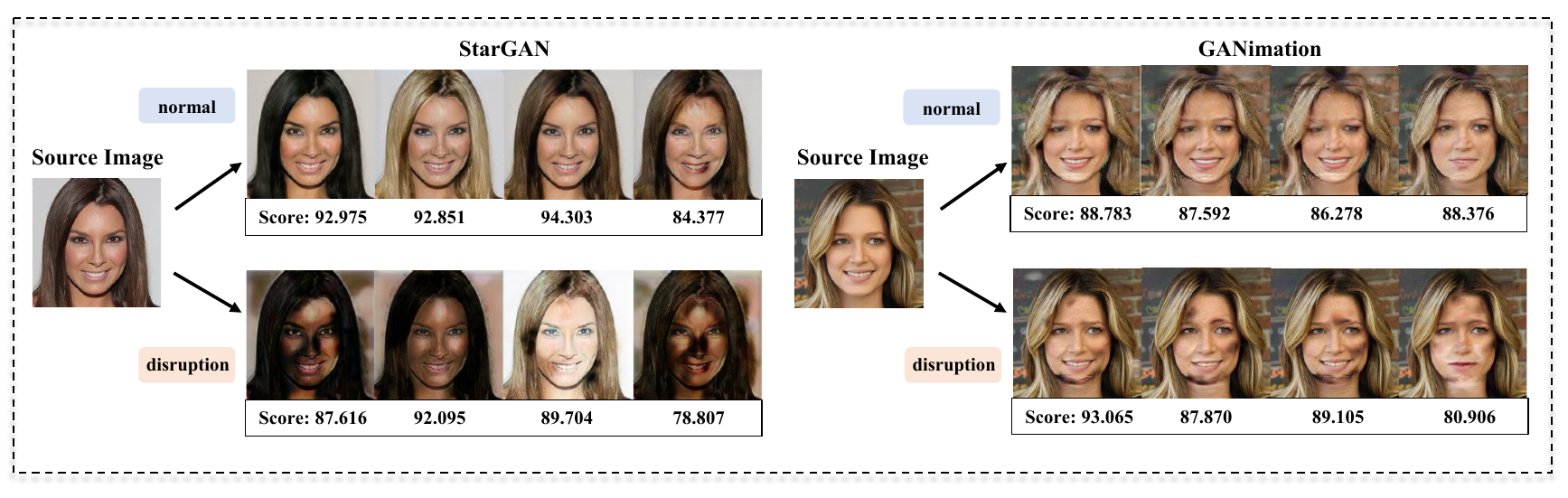} 
\caption{Visualization of Pixel-level Disruption. The "normal" row indicates images generated normally by the DeepFake models without any defense mechanisms. The "disruption" row indicates distorted images generated under protection using Ruiz \textit{et al.}'s disruption method. "Score" represents the similarity score between the generated images and the user's source images measured using the face recognition API provided by Face++, with a threshold of 69.101. If the score is greater than this threshold, it is considered to be the same identity.}
\label{Motivation}
\end{figure*}

\begin{figure*}[ht]
\centering
\vspace{10pt}
\hspace{-10pt}
{\includegraphics[width=3in]{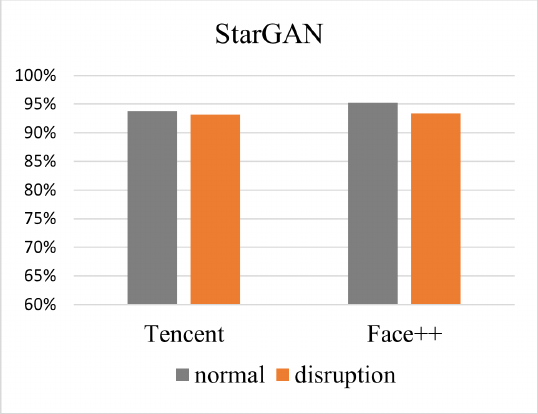}%
\label{stargan-white}}
\hfil
\hspace{-10pt}
{\includegraphics[width=3in]{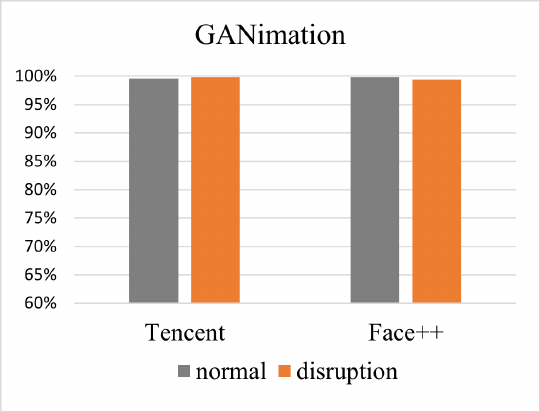}%
\label{ganimation-white}}
\hspace{-10pt}
\caption{FMRs on difference face verification APIs.}
\label{white}
\end{figure*}

\autoref{Motivation} and \autoref{white} illustrate qualitative visualizations and quantitative statistical results. From \autoref{Motivation}, it is evident that despite visually perceptible distortions in the generated images, they still have high confidence in matching the genuine identities of source images, with some distorted generated images even showing higher confidence than the original ones. 
Furthermore, \autoref{white} presents the evaluation results of FMR between distorted generated images and source images. Surprisingly, across different commercial facial recognition APIs, the FMR of the distorted generated images closely approximates that of the original generated images, reaching over 93\%. This indicates that even in a white-box scenario, pixel-level disruption methods are ineffective against face-swapping threats to identity authentication services. 

Therefore, it is imperative to disrupt the identity features of DeepFake samples, especially face-swapped images, to prevent the risks of identity theft and dissemination of false information.

\subsection{Description of Image Transformation Functions}
\label{A2}
When computing the protected images, we employ a differentiable random image transformation function to enhance the transferability and robustness against different face-swapping models. These transformations are described as follows:
1) Applies Gaussian blurring with random parameters. The kernel size is randomly chosen from \{3, 5, 7\}, and the sigma value ranges between 0.1 and 3.0.
2) Performs random cropping and resizing of the images. It generates cropped patches from the original image, applying a scaling factor between 0.25 and 4.

\subsection{Effectiveness of the Protected Images under Different Distance Budget}
\label{A3}
\begin{figure}[ht]
\centering
\begin{subfigure}{0.42\textwidth}  
  \centering
  \includegraphics[width=\linewidth]{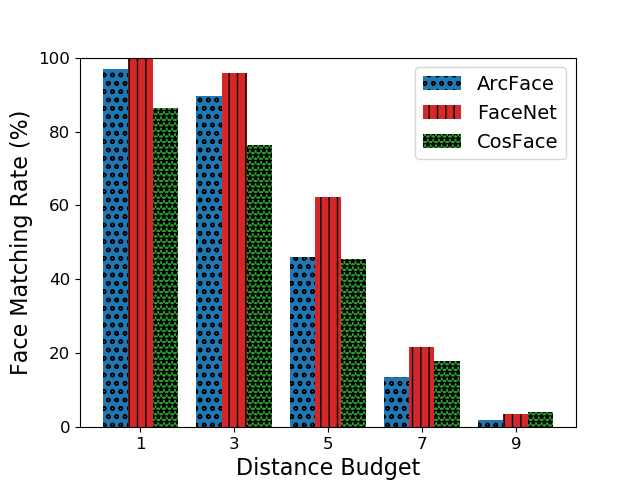}
  \caption{FaceShifter}
  \label{arcface_fs}
\end{subfigure}
\hspace{5pt}
\begin{subfigure}{0.42\textwidth}
  \centering
  \includegraphics[width=\linewidth]{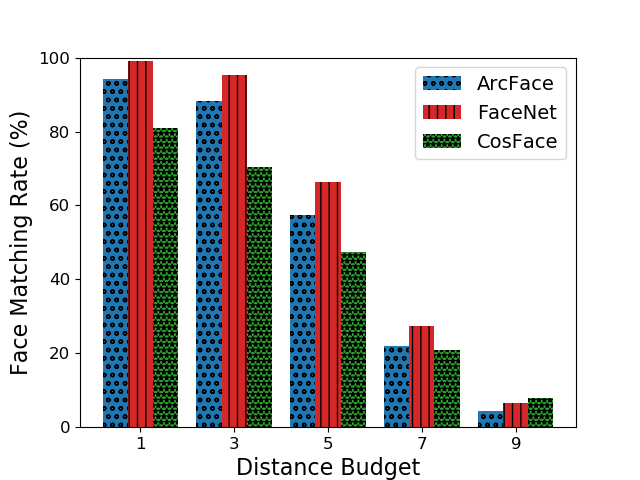}
  \caption{SimSwap}
  \label{arcface_ss}
\end{subfigure}
\vspace{-5pt} 
\caption{FMRs on face recognition models under different distance budgets (surrogate model: ArcFace)}
\label{black-eps-arcface}
\end{figure}

\begin{figure}[ht]
\centering
\begin{subfigure}{0.42\textwidth}  
  \centering
  \includegraphics[width=\linewidth]{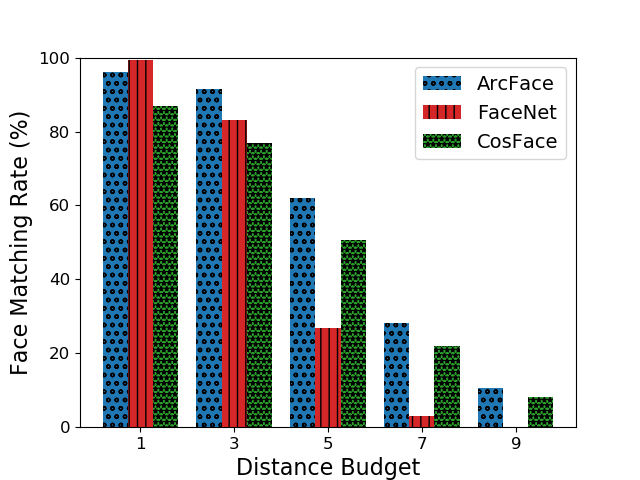}
  \caption{FaceShifter}
  \label{facenet_fs}
\end{subfigure}
\hspace{5pt}  
\begin{subfigure}{0.42\textwidth}
  \centering
  \includegraphics[width=\linewidth]{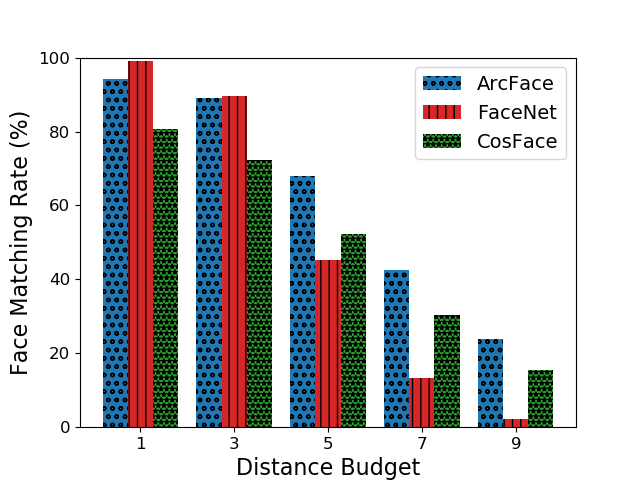}
  \caption{SimSwap}
  \label{facenet_ss}
\end{subfigure}  
\caption{FMRs on face recognition models under different distance budgets (surrogate model: FaceNet)}
\label{black-eps-facenet}
\end{figure}

We calculate protected images at different distance budgets $\epsilon$ on different surrogate models and then generate face-swapped images using the FaceShifter and SimSwap models. The evaluation results of the protected images obtained using the surrogate models ArcFace and FaceNet are presented in Figures \ref{black-eps-arcface} and \ref{black-eps-facenet}, respectively. We can see that FMRs decrease significantly as the distance budget $\epsilon$ increases in various cases. This suggests that protected images effectively mitigate the threats posed by face-swapping attacks to authentication and identification services reliant on facial recognition technology.

\subsection{Visualization Results of the Protected Images and Face-swapped Images}
\label{A4}
We present the visualizations of the protected images and corresponding face-swapped images based on the surrogate model ArcFace. Specifically, \autoref{black-adv-appendix} shows some visualization results of protected images calculated with our \texttt{FSG} under different distance budget $\epsilon$. In \autoref{black-gen-arcface-appendix} we provide the quantitative and qualitative evaluation results of face-swapped images generated by Faceshifter and SimSwap under the distance budget  $\epsilon=9$ . The evaluation results demonstrate the effectiveness of the protected images calculated on alternative Arcface in confusing visual perception.

\begin{table}[ht]
\caption{Visualizations of protected images at different budgets.}
    \centering
    \setlength{\tabcolsep}{1pt}
    \scalebox{0.9}
    {
    \begin{tabular}{c|c|c|c|c}
    \toprule[1pt]
       Source Image & $\epsilon=3$ &  $\epsilon=5$ &  $\epsilon=7$  & $\epsilon=9$\\
         \midrule
        \raisebox{-.5\height}{\includegraphics[width=0.19\linewidth]{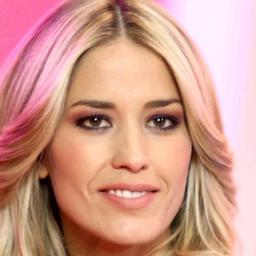}} & 
        \raisebox{-.5\height}{\includegraphics[width=0.19\linewidth]{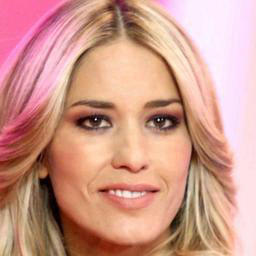}} & 
        \raisebox{-.5\height}{\includegraphics[width=0.19\linewidth]{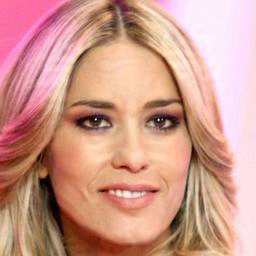}} &
        \raisebox{-.5\height}{\includegraphics[width=0.19\linewidth]{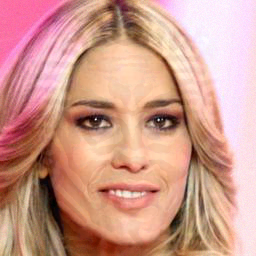}} &
        \raisebox{-.5\height}{\includegraphics[width=0.19\linewidth]{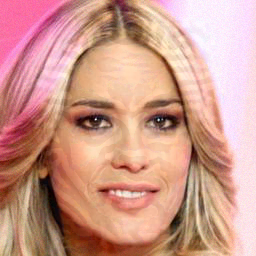}}
        \\
         \midrule
        \raisebox{-.5\height}{\includegraphics[width=0.19\linewidth]{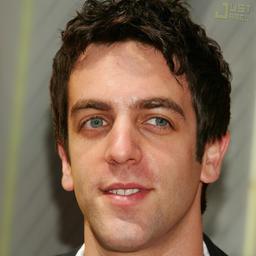}} & 
        \raisebox{-.5\height}{\includegraphics[width=0.19\linewidth]{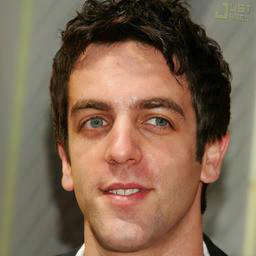}} & 
        \raisebox{-.5\height}{\includegraphics[width=0.19\linewidth]{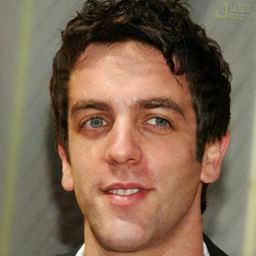}} &
        \raisebox{-.5\height}{\includegraphics[width=0.19\linewidth]{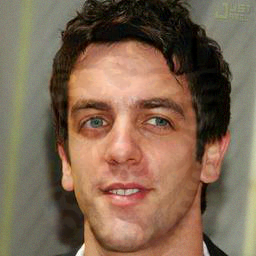}} &
        \raisebox{-.5\height}{\includegraphics[width=0.19\linewidth]{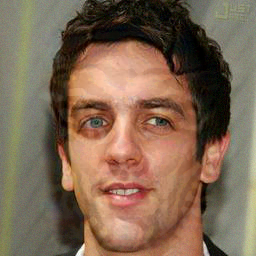}}
        \\
         \midrule
        \raisebox{-.5\height}{\includegraphics[width=0.19\linewidth]{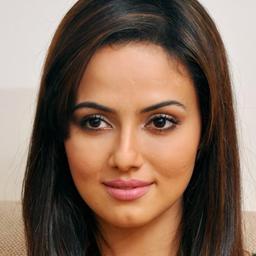}} & 
        \raisebox{-.5\height}{\includegraphics[width=0.19\linewidth]{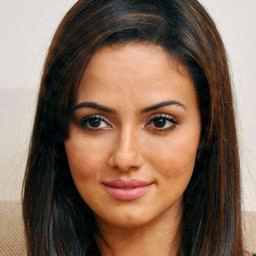}} & 
        \raisebox{-.5\height}{\includegraphics[width=0.19\linewidth]{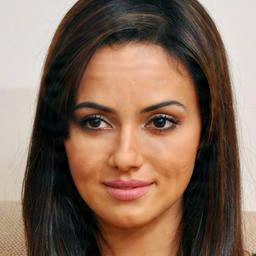}} &
        \raisebox{-.5\height}{\includegraphics[width=0.19\linewidth]{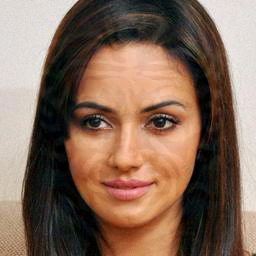}} &
        \raisebox{-.5\height}{\includegraphics[width=0.19\linewidth]{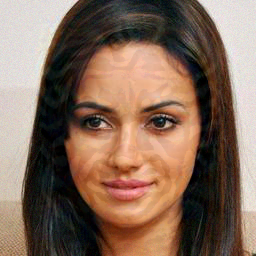}}
        \\
        \toprule[1pt]
    \end{tabular}}
    \label{black-adv-appendix}
\end{table}

\begin{table}[H]
  \caption{Quantitative and qualitative perceptual similarity of face-swapped images.}
    \centering
    \scalebox{0.9}
    {
    \begin{tabular}{c|c|c|c}
    \toprule[1pt]
       Source Image & Target Image &  FaceShifter &  SimSwap\\

         \midrule
        \raisebox{-.5\height}{\includegraphics[width=0.2\linewidth]{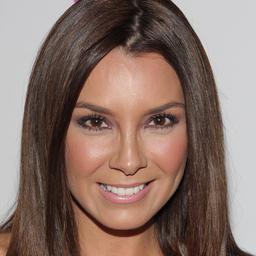}} & 
        \raisebox{-.5\height}{\includegraphics[width=0.2\linewidth]{./Evaluation/black/gen/target_000534.jpg}} & 
        \raisebox{-.5\height}{\includegraphics[width=0.2\linewidth]{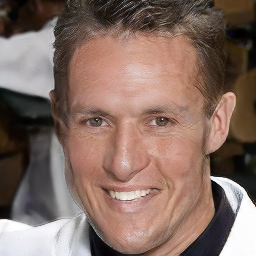}} &
        \raisebox{-.5\height}{\includegraphics[width=0.2\linewidth]{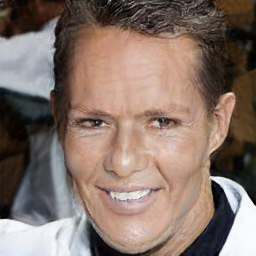}}
        \\
         \midrule
        \raisebox{-.5\height}{\includegraphics[width=0.2\linewidth]{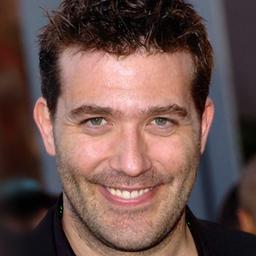}} & 
        \raisebox{-.5\height}{\includegraphics[width=0.2\linewidth]{./Evaluation/black/gen/target_000009.jpg}} & 
        \raisebox{-.5\height}{\includegraphics[width=0.2\linewidth]{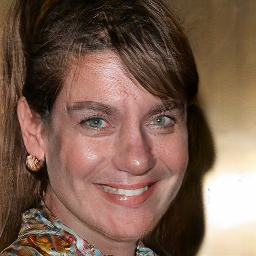}} &
        \raisebox{-.5\height}{\includegraphics[width=0.2\linewidth]{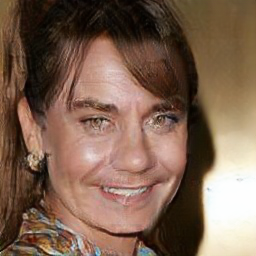}}
        \\
         \midrule
        \raisebox{-.5\height}{\includegraphics[width=0.2\linewidth]{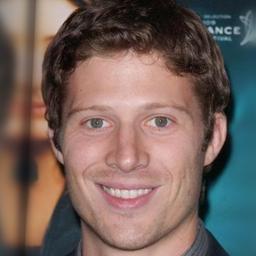}} & 
        \raisebox{-.5\height}{\includegraphics[width=0.2\linewidth]{./Evaluation/black/gen/target_000009.jpg}} & 
        \raisebox{-.5\height}{\includegraphics[width=0.2\linewidth]{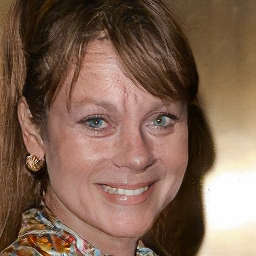}} &
        \raisebox{-.5\height}{\includegraphics[width=0.2\linewidth]{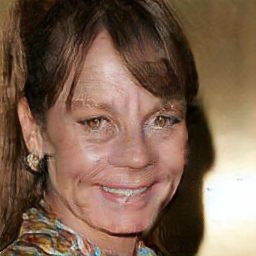}}
        \\
         \bottomrule
        \multicolumn{2}{c|} {LPIPS} &  0.5486 &  0.5397\\ 
        \toprule[1pt]
    \end{tabular}}
    \label{black-gen-arcface-appendix}
\end{table}

\subsection{Effectiveness on other DeepFake Models}
\label{A6}

\subsubsection{Evaluation on image-level face swapping}
We calculate the protected images based on different surrogate models and then generate corresponding face-swapped images using FSGAN, which involves facial reenactment and does not explicitly extract identity features. 
\autoref{FSGAN_result} shows the FMRs on various face recognition models and face verification APIs.

\begin{table}[htbp]
\centering
\caption{FMRs between images generated by FSGAN and the source images.}
\scalebox{1}{
\setlength{\tabcolsep}{3pt}
\begin{tabular}{c|l|lll}
\hline 
\multicolumn{2}{c}{}  & Baseline & ArcFace & Facenet \\ \hline 
\multirow{3}{*}{\begin{tabular}
[c]{@{}c@{}}Face recognition \\ Models\end{tabular}} 
& ArcFace & \multicolumn{1}{c}{73.9\%}   & \multicolumn{1}{c}{6.3\%}     & \multicolumn{1}{c}{13.5\%}    \\ 
& FaceNet & \multicolumn{1}{c}{75.9\%}   & \multicolumn{1}{c}{11.2\%}    & \multicolumn{1}{c}{ 0.3\%}  \\ 
& CosFace & \multicolumn{1}{c}{66.1\%}   & \multicolumn{1}{c}{17\%}      & \multicolumn{1}{c}{12.2\%}  \\ \hline \hline
\multirow{3}{*}{\begin{tabular}[c]{@{}c@{}}Face verification \\ APIs\end{tabular}}  
& Baidu  & \multicolumn{1}{c}{49.9\%}    & \multicolumn{1}{c}{0.3\%}     & \multicolumn{1}{c}{0.1\%}    \\ 
& Tencent & \multicolumn{1}{c}{66\%}   & \multicolumn{1}{c}{9.5\%}     & \multicolumn{1}{c}{4.1\%}   \\ 
& Face++  & \multicolumn{1}{c}{86.4\%}   & \multicolumn{1}{c}{21.5\%}    & \multicolumn{1}{c}{16\%}   \\ \hline 
\end{tabular}}
\label{FSGAN_result}
\end{table}

By comparing with the baseline results without any protection (in the 'Baseline' column), it is evident that the protected images generated on different surrogate models all significantly reduce the FMRs between the generated images and the source images. 
The effectiveness of protected images on FSGAN, including a face reenactment step, highlights the potential of our method to generalize to other DeepFake models (e.g., face reenactment techniques).
Additionally, we observe that the FMRs between the images normally generated by FSGAN without protection and the source images tend to be relatively low due to the lack of explicit constraints on identity information during generation. 
Compared with the state-of-the-art feature-level face-swapping methods FaceShifter and SimSwap, the identity preservation performance in images generated by FSGAN is inferior. 

\subsubsection{Evaluation on Diffusion-based Models}
Recent advancements in diffusion models have marked significant progress in image generation tasks ~\cite{kim2022diffusionclip}, showcasing higher-quality image synthesis compared to GANs. 
Preechakul \textit{et al.} ~\cite{preechakul2022diffusion}  utilize a learnable encoder to uncover high-level semantics, while employing diffusion probabilistic models (DPMs) as decoders to model the remaining stochastic variations, denoting this method as Diff-AE. 
We use Diff-AE to generate 500 pictures with modified attributes based on the protected images produced by \texttt{FSG}. 
The attributes include ``Bangs,'' ``Smiling,'' ``Eyeglasses,'' ``Mustache,'' and ``Pale\_Skin.''

\begin{figure*}[ht]
\centering
\begin{subfigure}{0.45\textwidth}  
  \centering
  \includegraphics[width=\linewidth]{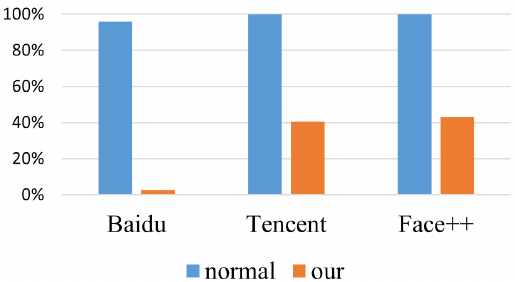}
  \caption{FaceShifter}
  \label{diffusion_defense}
\end{subfigure}
\begin{subfigure}{0.45\textwidth}
  \centering
  \includegraphics[width=\linewidth]{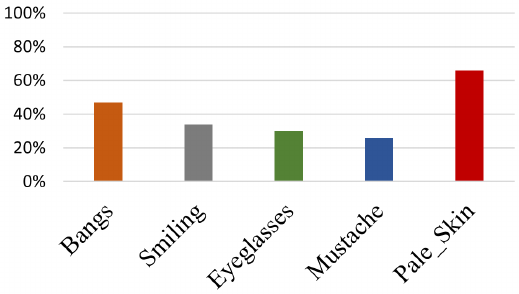}
  \caption{SimSwap}
  \label{diffusion_att}
\end{subfigure}
\caption{FMRs between images generated by Diff-AE and source images under different defenses and attributes.}
\label{diffusion}
\end{figure*}

As shown in \autoref{diffusion_defense},  without protection, the FMRs between images generated by Diff-AE and the user source images exceeds 95\%. However,  under our \texttt{FSG} protection, there is a varying degree of reduction in FMR across the three face verification APIs, particularly dropping to 2.8\% on Baidu. 
This indicates the potential for our method to transfer to other techniques under different scopes, despite the fact that the effectiveness of \texttt{FSG} across diffusion-based models is inevitably diminished by the limited transferability of adversarial perturbations on diverse network architectures.
Additionally, in \autoref{diffusion_att}, we further demonstrate the FMRs between different types of attribute-modified images and the source images under the protection of our \texttt{FSG}. It can be seen that the effectiveness of \texttt{FSG} varies across different attributes, with notable effectiveness observed on all attributes except for ``Pale\_Skin.'' 
Future endeavors will delve deeper into this phenomenon.

\end{document}